\documentclass{article}

\usepackage{arxiv}

\usepackage[utf8]{inputenc} %
\usepackage[T1]{fontenc}    %
\usepackage{hyperref}       %
\usepackage{url}            %
\usepackage{booktabs}       %
\usepackage{amsfonts}       %
\usepackage{nicefrac}       %
\usepackage{microtype}      %
\usepackage{lipsum}

\usepackage{amsmath}    %
\usepackage{subcaption} %
\usepackage{graphicx}   %
\usepackage{courier}    %
\usepackage{graphicx}
\usepackage{bm}
\usepackage{algorithm}
\usepackage{algorithmicx}
\usepackage{algpseudocode}
\usepackage{caption}
\usepackage{subcaption}
\usepackage{tabularx}
\usepackage{enumitem}
\usepackage{makecell}
\usepackage{xcolor}     %

\begin{document}

\title{Recommending POIs for Tourists by User Behavior Modeling and Pseudo-Rating}

\author{
  Kun Yi\\
  \And
  Ryu Yamagishi\\
  \And
  Taishan Li\\
  \And
  Zhengyang Bai\\
  \And
  Qiang Ma\\
}

\maketitle

\begin{abstract}
POI recommendation is a key task in tourism information systems. However, in contrast to conventional point of interest (POI) recommender systems, the available data is extremely sparse; most tourist visit a few sightseeing spots once and most of these spots  have no check-in data from new tourists. Most conventional systems rank sightseeing spots based on their popularity, reputations, and category-based similarities with users' preferences. 
They do not clarify what users can experience in these spots, which makes it difficult to meet diverse tourism needs. 
To this end, in this work, we propose a mechanism to recommend POIs to tourists. Our mechanism include two components: one is a probabilistic model that reveals the user behaviors in tourism; the other is a pseudo rating mechanism to handle the cold-start issue in POIs recommendations. We carried out extensive experiments with two datasets collected from Flickr. The experimental results demonstrate that our methods are superior to the state-of-the-art methods in both the recommendation performances (precision, recall and F-measure) and fairness. The experimental results also validate the robustness of the proposed methods, i.e., our methods can handle well the issue of data sparsity.       
\end{abstract}


\section{Introduction} \label{introduction}
Tourism is a major industry in the world economy and plays an important role in our lives. Automated discovery and recommendation of sightseeing spots are drawing more attention to meet the growing demands for personalized tours \cite{zhuang2017sns,DBLP:journals/mta/GeZM19,DBLP:journals/ijbdi/ShenGZM18,lim2015personalized,zhou2017places}. In recent years, social network services (SNS) have been spreading extensively, and the Internet is filled with various travel-related information. 
Massive amounts of data, such as geo-tagged images and videos, have been posted to SNSs such as Flickr and Instagram to share sightseeing experiences. These geo-tagged images are widely used in landmark recognition and tour recommendations \cite{10.1007/s11042-010-0623-y}.

Point-of-interest (POI) recommendation is a key challenge in location-based social networks (LSBNs)\cite{DBLP:journals/mta/ZhengZC11,islam2020survey,yuan2013and, yuan2015and}. The information generated and shared by users on LSBNs is useful for understanding user preferences for POI recommendations. Generally, POI recommendations aim to predict or recommend the next set of POIs in which a user may be interested by learning from the check-in history and related information of that user. To this end, Markov chain-based approaches, matrix factorization (MF) methods, and deep learning-based methods have been proposed \cite{ference2013location,lian2014geomf,lian2018geomf++,DBLP:journals/mta/ZhengZC11,islam2020survey,DBLP:journals/www/XuFCLW20}.             

Similar to conventional POI recommendation tasks \cite{DBLP:journals/kais/LimCKL19}, tourists tastes must be captured from their behaviors for sightseeing recommendations. However,  
unlike daily activities, the movements of users when sightseeing are not regular and difficult to predict.
We visit the same places many times in our daily lives, such as home, school, and workspaces, and often repeat the identical things every day.
However, when acting as tourists, we rarely go to the same place many times or repeat the same activities as in our daily lives. When we travel, we are challenging, adventurous, and explore things that we rarely experience in our daily lives. This results in different behavior patterns. The existing MF-based methods (such as GeoMF++\cite{lian2018geomf++,lian2014geomf} and LGLMF\cite{Rahmani2020LGLMF}) and deep-learning-based methods (SAENAD\cite{Ma2018SAENAD},etc.) could not handle this problem well. Moreover, for a tourist, there is little or no check-in data at the destination city because it is likely their first visit the city. A key challenge in sightseeing recommendation is modeling tourists' behaviors with extremely sparse data, which is more difficult than modeling daily life recommendations. In this sense, sightseeing spot recommendation is one of the hardest recommendation tasks. We define a new class of POI recommendation tasks, namely sightseeing spot recommendations, which focuses on the challenges of little or no history data and different user behavior patterns (Section \ref{sec:pre}). 

Recommender systems that analyze tourists' travel behaviors are receiving  significant attention \cite{ference2013location, lian2014geomf, lian2018geomf++, lim2015personalized, lim2016towards, yin2016joint, yin2013lcars, yin2015joint}. Most existing studies focused on spot popularity and the relationship between users and locations, which infer users' preferences and opinions to recommend personalized tours and sightseeing spots \cite{ference2013location,liu2013personalized,sun2020dexa,zhuang2017sns,DBLP:journals/kais/LimCKL19}. The popularity of a POI is based on the check-in numbers, stay duration, and so on. It is a reasonable factor in tourism because tourists are unfamiliar with the destination city and do not have historical check-in data. However, the popularity bias is the main disadvantage: popular locations are always being recommended while the majority of other locations do not get the deserved attention. It cannot recommend locations with sufficient consideration of what the user want and can do at the spots. On the other hand, fairness is an increasingly important area in recommender system \cite{abdollahpouri2019unfairness, ashkan2014diversified, steck2018calibrated}. The popularity based recommendation could not meet the diverse needs of tourists, especially the ones who want to enjoy deep and personalized tourism, which results in lack of fairness among tourists.
To meet diverse tourists' needs, some researchers proposed methods that reveal ``what to see" for sightseeing spot  recommendation, based on the estimation of tourist attractions and focusing on the scenery and landscape \cite{DBLP:journals/mta/GeZM19,DBLP:journals/ijbdi/ShenGZM18}. In this work, we propose a novel user-experience-oriented sightseeing recommender mechanism that focuses on sightseeing activities other than ``seeing", such as ``eating", ``playing", and ``shopping", and makes user adaptive recommendations based on  ``what tourists want to do or enjoy there". 

We propose a Basic User Experience Model (B-UEM), which is a probabilistic generation model for analyzing tourist behaviors in sightseeing spots. 
Inspired by \cite{yuan2013and, yuan2015and}, the proposed user experience model involves four factors: Who (does it), What (the tourist does), Where (the tourist does it), and When (the tourist does it). With B-UEM, we reveal ``what we can do/enjoy there" and recommend sightseeing resources based on user behaviors.  
A pseudo rating mechanism is also proposed to handle cold-start scenario, in which tourists are new to the city and have no historical data. Several keywords about Where and What are provided to new tourists to be rated as start-up information and recommendations are made based on B-UEM.

In short, we recommend sightseeing spots considering ``what we can experience there" and ``what we want to experience there" rather than just ``where we can or should visit".
The major contributions of this work are summarized as follows.
\begin{itemize}
 \item We propose a probabilistic model to analyze user experiences using geo-tagged images. We model the user experience by introducing a hidden factor (Section \ref{proposed_method}). Conventional models\cite{yuan2013and,yuan2015and} generate time, words, and location for each post (tweet, etc.) by a user. Such approaches require each user has enough records to learn the model. This fashion makes it not suitable for modeling tourist behaviors. In contrast, our user experience model generates users, locations, and words per each post (image, etc.) from the hidden user experiences. This feature makes our model suitable for mitigating the sparse data issue in tourism recommender systems. In addition, our experimental results also reveal the differences between modeling user behaviors for daily lives and for tourism (Section \ref{model_comparison}).
 \item We propose a novel approach to handle the cold-start problem, especially for new tourists without historical data (Section \ref{model_application}). In addition to the user behavior modeling, we propose a novel pseudo rating mechanism to recommend sightseeing spots to new tourists with considering their preferences.  
 \item Extensive experiments with tourist data from two real-world datasets, YFCC100M Kyoto \cite{sun2020dexa} and User Visits \cite{lim2015personalized}, reveal that the proposed methods are superior to other state-of-the-art approaches (Section \ref{experiments}). The proposed methods not only achieve higher precision, recall and F-measure, but also avoid popularity bias to achieve satisfaction of more users and make recommendation more diversely at the same time. 
\end{itemize}

\section{Related Work} \label{related_work}

Generally, people only visit a small number of POIs, and the related check-in data is sparse. This feature makes POI recommendation harder than typical recommender systems, such as product recommendation in electronic commerce systems. In the field of POI mining and recommendation, it is important to model the user behaviors. Song et al. claimed that it was possible to predict user behaviors with up to 93\% accuracy by measuring the amount of information in the trajectory of daily living behaviors \cite{song2010limits}. A general temporal pattern was developed for user location prediction from a user's check-in behaviors \cite{gao2013exploring}\cite{gao2013modeling}. They also stated that behavior during travel is very different. Other researchers tackled the sparse data issue and proposed models that could handle in and out of town behavior \cite{ference2013location, wang2015geo, yin2016joint, yin2013lcars, liu2017experimental}. Similar to conventional POI recommendations, the check-in data in tourism is also sparse because that a tourist only visits a few spots (POIs). In addition,  
one of the challenges in predicting tourist behaviors is the absence of reproducibility, even for the same tourist. In tourism, people rarely go to the same place repeatedly.
It is common for a traveler to go to a particular spot and never re-visit there again, making the relationships between a tourist and sightseeing spots extremely sparse. Therefore, more efficient approaches for tourism recommendations are necessary. 
In addition to the problem of data sparsity, the concept of groups was introduced to address the cold start issue \cite{wang2015geo,yin2016adapting,zhuang2017sns}.
In conventional research, groups are classified using only the features of users or items. In our work, users, locations, and behaviors are comprehensively analyzed and classified into hidden factors that represent user experiences in tourism. 

There are also several methods that focus on user behavior mining \cite{aliannejadi2018personalized,rahimi2019behavior,yin2015joint}.
In these methods, the information on user behaviors and locations is used to predict the location. However, this is insufficient for capturing the user's interests and preferences.
Specifically, probabilistic models of $W^4$ and $EW^4$ to model users’ behaviors on Twitter have been proposed \cite{yuan2013and,yuan2015and}.
Latent Dirichlet allocation was used to model locations and behaviors while ignoring the influences of time and users \cite{wang2007mining}. 
Geographic topics from Twitter users were utilized to analyze the Who, Where, and What using a probabilistic model \cite{hong2012discovering}.
A probabilistic model was also used for user behavior prediction \cite{wang2018tpm,yin2016discovering}.
In contrast to our work, the target user's behaviors in these methods are on daily activities rather than sightseeing. Users’ behaviors during sightseeing are different from those in daily lives, and thus a different model is necessary. Furthermore, historical user data in the destination city is extremely sparse when compared with conventional POI recommendations. To handle this challenge, our user experience model generates users, locations, and words from the hidden user experiences. This differs from conventional models that generate time, words, and location for each post (tweet, etc.) from a user.   

Tweets were often utilized as user behavior data \cite{yuan2013and,yuan2015and}. In contrast, we use images from YFCC100M \cite{thomee2016yfcc100m} as the dataset to train our user-experience model.
The main difference between Twitter and Flickr or Instagram is their dominant data types, text or images. 
Images have more information than freely written text data and contain information about the user's tastes and behaviors.
Text data are sometimes composed of emotional, subjective, and temporary elements. It is possible to obtain more accurate and objective information on user behavior by analyzing images rather than text data. Images reflect users’ tastes and preferences more accurately than check-in data, such as the data from Foursquare and Gowalla. Previous researchers have attempted to predict places from image datasets such as Flickr and Instagram \cite{li2015you,liu2014your,wang2017your}. However, they did not consider the data sparseness and semantic meaning of images, and determined users’ preferences directly from the data.

\section{User Experience Model} \label{proposed_method}
\subsection{Preliminary} \label{sec:pre}
In this study, we denote the sets of tourists and sightseeing spots in a sightseeing city $c$ as $U$ and $L$, respectively. The tourist records of $U$ in city $c$ are denoted by $D$. Table \ref{variable_definition} illustrates the primary notations used in this study.

\begin{description}[leftmargin=0cm]
 \setlength{\leftskip}{0cm}
\item[\textbf{Sightseeing Spot}]
A sightseeing spot $l$ is geo-coded by a coordinate $(x_p,y_p)$, where $x_p$ and $y_p$ denote the latitude and longitude, respectively. In a sightseeing recommender system, the destination sightseeing spots $L = \{l_1,...l_m\}$ are pre-defined by the system or automatically detected by using the historical tourist records $D$. Generally, a sightseeing spot is a specific coordinate that attracts tourists and is usually represented as a point or coordinate on a map. 

\item[\textbf{User Experience}]
A user experience $e$ is a tuple of $(u, l, t, W)$ and denotes the user $u$ (\textit{who} is the tourist) engaging in behaviors $W$ (\textit{what} did $u$ do) at time $t$ (\textit{when} did $u$ visit there) in spot $l$ (\textit{where} did $u$ visit). We use words that represent activities and scenes to express user behaviors in sightseeing spots. 

\item[\textbf{Tourist Records}]
The tourist records $D$ are collected in the given city $c$ from tourists $U$. Without losing generality, we use the geo-tagged images uploaded on social image sharing sites, such as Flickr and Instagram. Here, a geo-tagged image $d$ is denoted as $(u_d,l_d,t_d,W_d)$, which signifies that $u_d$ took a photo representing user behaviors $W_d$ at time $t_d$ in location $l_d$. A geo-tagged image (a tourist record)  represents one or more user experiences. Therefore, $D$ is a set of $d$ from tourist $U$ who visited and took photos in sightseeing spots $L$ in $c$.  Existing methods are applied to extract words describing user behaviors \cite{zhou2017places}.   

\item[\textbf{Sightseeing Spot Recommendation}]
Sightseeing spot recommendation is a class of POI recommendation tasks that recommends sightseeing spots $L=\{l_1,...l_m\}$ in city $c$ for a tourist $u'$, given tourist records $D$ in $c$ of tourists $U$. Note that in general, $u' \notin U$, which differs from conventional POI recommendations.    

\end{description}

\subsection{Assumptions} \label{assumption}
This study focuses on recommending spots based on what tourists want to experience at that location. To this end, we first propose a novel user experience model to analyze tourist behaviors in the destination city $c$ by considering the following assumptions. 

\begin{description}[leftmargin=0cm]
  \setlength{\leftskip}{0cm}
  \item[\textbf{Assumption 1:}] Tourists may enjoy similar experiences and activities. In other words, we can group and generate users based on these activities and experiences. Conversely, an experience-based user group contains multiple tourists with similar preferences. Obviously, a tourist may enjoy  different experiences.  
  \item[\textbf{ Assumption 2:}] Each spot serves more than one experience, and similar spots may provide similar sightseeing experiences. In other words, we can group and generate spots based on user experiences. An experience could be enjoyed in several spots. 
  \item[\textbf{ Assumption 3:}] Tourists enjoy different experiences and engage in different activities at different spots. An experience represents a group of similar activities or actions performed by tourists at their visited spots. For example, taking pictures, eating sweetmeats, and doing Zen meditation are common actions in sightseeing spots. We may group actions and behaviors based on user experiences. 
  \item[\textbf{ Assumption 4:}] User experiences are influenced by the time when the tourists visit spots. In other words, visiting time may be grouped based on user experiences. 
\end{description}
  \begin{table}[!tb]
  \begin{center}
  \caption{Symbols}  
  \label{variable_definition}
  
  \begin{tabularx}{0.5\textwidth}{ >{\hsize=.3\hsize\centering\arraybackslash}X | >{\hsize=1.7\hsize\centering\arraybackslash}X }
    \hline
    Symbol & Description \\
    \hline
    \bm{$G$} & Set of User Experiences \\
    \bm{$U$} & Set of Tourists  \\
    \bm{$D$} & Set of Tourist Records \\
    \bm{$L$} & Set of Sightseeing Spots \\
    \bm{$W$} & Set of Words Describing Users' Activities  \\
    \bm{$T$} & Set of Time Slots \\
    $N_w$ & Number of Words in Data $d(\in D)$ \\
    \hline
    $d$ & Tourist Record ($\in D$) \\
    $g_d$ & User Experience to which $d$ belongs  \\
    $u_d$ & User (tourist) who created $d$ \\
    $l_d$ & Spot where $d$ was created \\
    $w_d$ & Words ($\in W$) appearing in $d$ \\
    $t_d$ & Time when $d$ was created \\
    \hline
    $\theta$ &  Categorical Distribution over $G$ \\
    $\rho_{t_d}$ & Categorical Distribution over $W$ Specific to $t_d$ \\
    $\mu_{l_d}$ & Categorical Distribution over $W$ Specific to $l_d$ \\
    \hline
    $\pi_{g_d}$,$\phi_{g_d}$, & Categorical Distribution over $U$, $L$, \\
    $\sigma_{g_d}$,$\tau_{g_d}$ & $W$ and $T$ Specific to $g_d$ \\
    \hline
    $\alpha, \beta, \gamma$, $\delta$, $\epsilon$, $\eta$, $\iota$, $\kappa$ & Dirichlet Priors to Categorical Distributions \\
    \hline
  \end{tabularx}
  \end{center}
  \end{table}

\subsection{User Experience Model} \label{model_definition}
\begin{figure}[!tb]
 \begin{center}
    \includegraphics[scale=0.18]{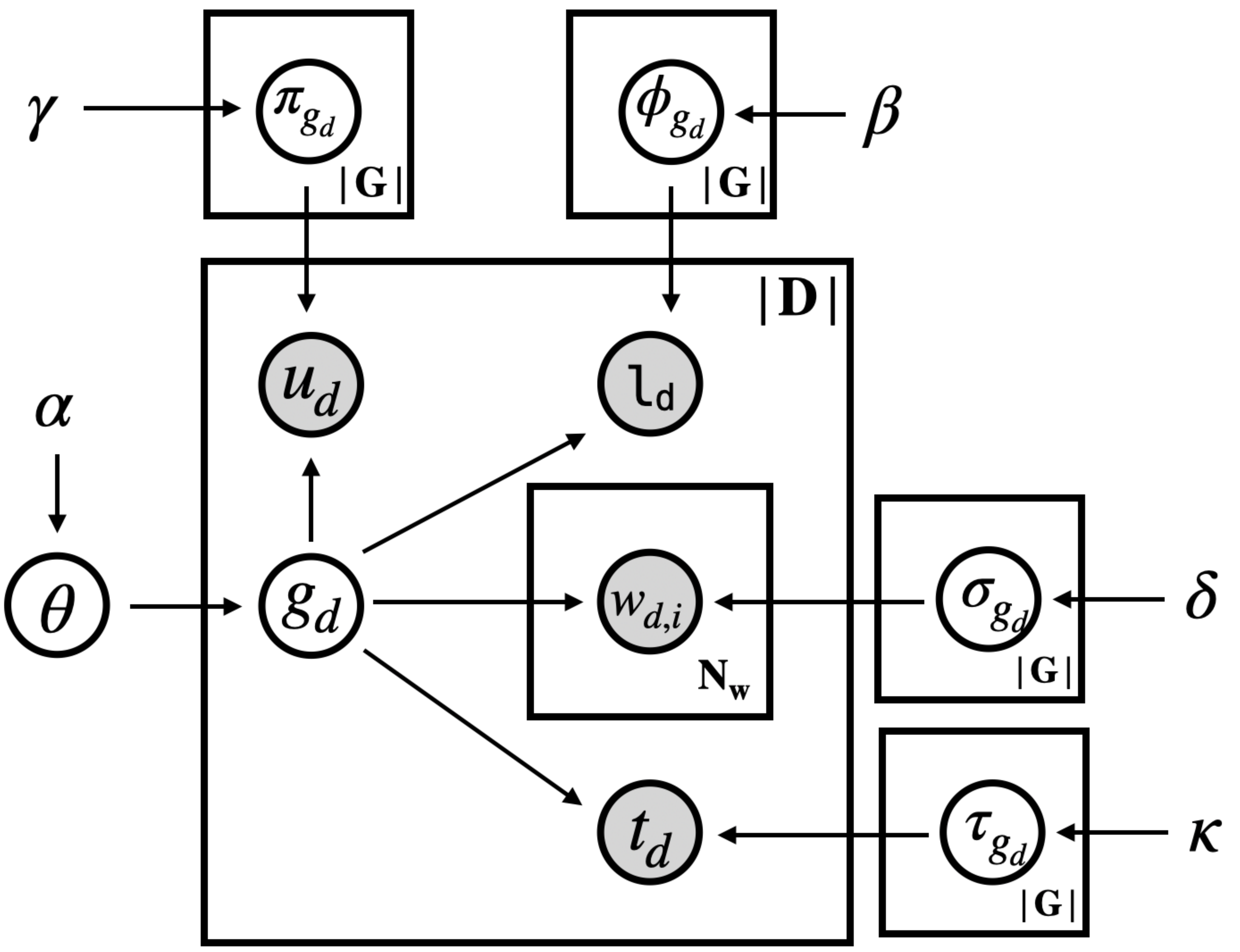}
    \caption{Graphical model of the B-UEM. }
    \label{basemodel}
 \end{center}
\end{figure}

We construct a basic user experience model (B-UEM) to analyze tourists’ behaviors, considering Assumptions 1, 2, 3, and 4.
Tourists, spots, times, and behaviors are treated as observed variables, and the user experience is considered as a latent variable. As previously mentioned, our user experience model generates tourists, spots, and behavior words using the hidden user experience to address the extreme sparsity of tourist data. The B-UEM graphical model is illustrated in Figure \ref{basemodel}. 
The generative process in the B-UEM model is shown in Algorithm \ref{base_generative_part}. User experience $g_d$ is drawn from a categorical distribution with parameter $\theta$. Then, using $g_d$, we generate tourist $u_d$, spot $l_d$, and a set of words $w_{d,i}$ ($i=1,...,N_w$) related to $g_d$. Note that $w_{d,i}$ in the graphical model indicates the $i$th word among the words indicating the behavior at a certain tourist record $d(\in D)$. The white circles in the graphical model represent the latent variables, and the circles with a gray background color represent the observed variables (see Algorithm \ref{base_generative_part}).

One of the notable features of our model is that it generates tourists, spots, and behavior words from the user experience, a latent factor. As a result, the four factors, $u, l, t$, and $w$, are conditionally independent.
Each $w$ is influenced only by the user experience. This guarantees that they are independent and identically distributed without influencing one other.

\begin{algorithm}[!tb]
  \caption{Generative Process of the B-UEM Model}\label{base_generative_part}
  \begin{algorithmic}[1]
    \State $\theta \sim Dir(\alpha)$ \Comment{draw priors to user experiences}
    \For{$g$ in $|\bm{G}|$}
    \State $\pi_{g} \sim Dir(\gamma)$ \Comment{draw priors to tourists}
    \State $\phi_{g} \sim Dir(\beta)$ \Comment{draw priors to spots}
    \State $\sigma_{g} \sim Dir(\delta)$ \Comment{draw priors to words describing behaviors}
    \State $\tau_{g}  \sim  Dir(\kappa)$ \Comment{draw priors to time slots}
    \EndFor
    \For{$d$ in $|\bm{D}|$} \Comment{proceed in data}
    \State $g_d \sim Categorical(\theta)$ \Comment{draw a user-experience}
    \State $u_d \sim Categorical(\pi_{g_d})$ \Comment{draw a tourist using user-experience}
    \State $l_d \sim Categorical(\phi_{g_d})$ \Comment{draw a spot using user-experience}
    \State $t_d  \sim  Categorical(\tau_{g_d})$ \Comment{draw a time using user-experience}
    \For{$i$ in $N_w$}
    \State $w_{d, i} \sim Categorical(\sigma_{g_d})$ \Comment{draw words using user-experience}
    \EndFor
    \EndFor
  \end{algorithmic}
\end{algorithm}

Tourist record $d$ consists of tourist $u_d$, visited spot $l_d$, time $t_d$, and activities $w_d=\{w_{d, 0},\cdots ,w_{d, N_w}\}$ performed there.
These are generated from corresponding categorical distributions and represented using discrete distributions. 
Tourists, spots, words, and times are generated by the hidden user experience $g$. 
$g_d$ is generated from the categorical probability distribution with parameter $\theta$, denoted as
\begin{equation}
  g_d \sim Categorical(\theta).
\end{equation}

From Assumption 1, a user could be generated by a user experience. Hence, the following equation applies.
\begin{align}
  u_d \sim & Categorical(\pi_{g_d}) \label{user_generation} \\
  \pi_{g_d} \sim & Dirichlet(\alpha) \nonumber
\end{align}
Similarly, spots are generated based on Assumption 2 from the probability distribution $\phi_{g_d}$ of spots for each user experience.
\begin{align}
  l_d \sim & Categorical(\phi_{g_d}) \label{location_generation} \\
  \phi_{g_d} \sim & Dirichlet(\beta) \nonumber
\end{align}
Time slots are also generated from the probability distribution $\tau_{g_d}$ by a user experience according to Assumption 4:
\begin{align}
  t_d \sim & Categorical(\tau_{g_d}) \label{time_generation} \\
  \tau_{g_d} \sim & Dirichlet(\kappa) \nonumber
\end{align}
Finally, words describing behaviors are denoted by $w_{d, i}$, and the fact that this variable stands alone indicates that there are multiple ($N_w$) actions for a tourist record whose tourist and spot are determined (Assumption 3).
For each word, generation is performed as follows:
\begin{align}
  w_{d, i} \sim & Categorical(\sigma_{g_d}) \\
  \sigma_{g_d} \sim & Dirichlet(\delta) \nonumber
\end{align}
%
\subsection{Recommending Spots with Pseudo Rating Mechanism} \label{model_application} 
The proposed B-UEM model involves four aspects of tourist behaviors: who, where, when, and what. Similar to previous user behavior models, missing information can be inferred in some of these aspects by using other given aspects. 
Here, without losing generality, to a tourist $u \in U$ whose past records have been known, we demonstrate the spot recommendation with $p(l|u)$. For location $l_i \in L$, we have   
\begin{align}
  p(l_i | u) \propto &  \sum_{g_m}p(l_i | g_m,u)p(g_m|u) \nonumber \\
  = & \sum_{g_m}p(l_i | g_m)p(g_m|u) \\
    p(g_m|u) \propto & p(u|g_m)p(g_m) 
\end{align}
\noindent Hence, $p(l_i|u)$ is calculated as follows:
\begin{align}
  p(l_i|u) 
  = & \sum_{g_m \in G} \phi_{g_m}(l_i) \pi_{g_m}(u) \theta(g_m)
\end{align}

Similarly, we can recommend activities by estimating $p(w_j|u)$ as follows: 
\begin{align}
     p(w_j|u) 
  =& \sum_{g_m \in G} \sigma_{g_m}(w_j) \pi_{g_m}(u) \theta(g_m)
\end{align}

\paragraph{Pseudo Rating Mechanism}
In contrast to the conventional recommendation systems, the challenge here is that there is no data for a new tourist $u'$. Therefore, our model cannot be directly applied to recommend sightseeing resources to $u'$. To handle this challenge, we propose a pseudo rating mechanism. Suppose that $n$ images from totally $N(n<<N)$ tourism records of other users are provided to a new user $u'$, and $u'$ rates the images. Under the assumption that a user's rating scores of all images $\{r_i\}_{i=1}^N$ follow a known distribution $F_{score}$, the rating scores can be inferred and used to predict the spots and behaviors that $u'$ may prefer.

We denote the rating information by $R = \{ (r_i,l_i,w_i, I_i)\}_{i=1}^n$, where $l_i, w_i$ are the location and activity of the image $I_i$, respectively. By law of total probability,
\[ p(l,w|R) = \sum_{g \in G} p(l,w|u' \in g,R)p(u' \in g|R) \]
As we suppose that $R$  is fully determined by $u'$ while deciding the preference, $P(l,w|u' \in g,R) = P(l,w|u' \in g)$, which means
\[ p(l,w|R) = \sum_{g \in G} p(l,w|u' \in g)p(u' \in g|R)\]
Apply Bayes' theorem, 
\begin{align}
    p(u' \in g|R) &= \frac{p(R|u' \in g)p(u' \in g)}{p(R)} \\
    \propto& \prod_{i=1}^{n}p(scores(l_i, w_i, I_i)=r_i|u' \in g)p(u' \in g)
\end{align} 
$P(u' \in g)$ can be calculated easily with the posterior, and under our assumption of rating score distribution, 
\begin{align}
&p(scores(l_i, w_i, I_i)=r_i|u' \in g) \\
&=  p\Big(\frac{N +1- rank(l_i, w_i, I_i)}{N} = F_{score}(r_i)\Big|u' \in g\Big) 
\end{align}
where, $rank(l_i, w_i, I_i)$ is the rank of $(l_i, w_i, I_i)$ in $N$ images. However, the accurate calculation of the distribution of $rank(l_i, w_i, I_i)$ has the same time complexity as matrix inversion, therefore we have to approximate the distribution by Monte Carlo sampling.

Now, we have two UEM-based methods to recommend location to deal with different situations.
\begin{itemize}
    \item To users whose records have been known and used as train data, we could recommend location based on UEM without pseudo rating mechanism.
    \item To new users whose records remain unknown, we could provide several images(or locations, words describing activities) to be rated, and make recommendation based on pseudo rating mechanism and UEM.
\end{itemize}
According to different type information provided to new users, we classify the second method into three sub-methods, namely, al(locations provided), wtol(words describing activities provided) and wl(both words and locations provided).

In addition, if we could receive a feedback of recommendation,  $p(u' \in g|R)$ could be used as prior probability for next recommendation, which hence forms a user-adaptive recommender system for new tourists. 
\section{Experimental Evaluation} \label{experiments}
In this section, first, we evaluated the performance of the proposed model for modeling tourist behaviors against its variations. Then, we compared the recommendation methods based on our model against three state-of-the-art methods, GeoMF++ \cite{lian2018geomf++}, SAE-NAD \cite{Ma2018SAENAD} and LGLMF \cite{Rahmani2020LGLMF},for spot recommendation. The datasets and the source code can be found at https://github.com/Ma-Lab-Public/UX-SPR.

The model was trained with Stochastic Variational Inference \cite{bingham2019pyro} implemented using Pyro. Adam \cite{kingma2014adam} was used as the optimization method, and its learning rate was set to $0.001$.
\subsection{Dataset} \label{dataset}
Two real-world datasets were used: YFCC100M Kyoto \cite{sun2020dexa} and User Visits \cite{lim2015personalized}\cite{lim2016towards}.  
Both datasets were collected from Flickr. For YFCC100M Kyoto, we detected the spots by applying the city adaptive clustering method \cite{sun2020dexa}. The User Visits dataset consists of geo-tagged images from eight cities. 
To filter noisy data for all datasets, we remove users who visited less than 2 POIs. We also discretized the time in months. The dataset statistics after preprocessing are shown in Table \ref{dataset-table}.

\begin{table}[!bt]
  \begin{center}
    \caption{Dataset Statistics}
    \label{dataset-table}
    \begin{tabular}{ c|c|c|c } \hline \hline
      City & \makecell[c]{\# of Users\\(Tourists)} & \# of Spots & \makecell[c]{\# of Images\\(Tourist Records)} \\  \hline
      Kyoto & 783 & 1496 & 55357 \\ 
      Budapest & 547 & 38 & 15911 \\
      Delhi & 119 & 23 & 2460 \\
      Edinburgh & 751 & 28 & 26658 \\
      Glasgow & 291 & 27 & 9031 \\
      Osaka & 186 & 27 & 6153 \\
      Perth & 65 & 22 & 2835 \\
      Toronto & 650 & 29 & 28087 \\
      Vienna & 573 & 28 & 27363 \\ \hline \hline
    \end{tabular}
  \end{center}
\end{table}

As previously mentioned, the words that indicate behaviors were extracted from images using two scene detection methods, place\cite{zhou2017places} and attribute\cite{patterson2014sun}, and for each image we used the top twelve words when using place\cite{zhou2017places} and the top four words when using attribute\cite{patterson2014sun} based on the pre-experimental results. 
\subsection{Model Comparison}\label{model_comparison}
We compare the proposed B-UEM model with its variations. To define the comparative models, we introduce two more assumptions that are standardly applied for user behavior modeling. 
\begin{description}[leftmargin=0cm]
  \setlength{\leftskip}{0cm}
 \item[\textbf{ Assumption 5:}] The behaviors at a spot are influenced by the location (spot), i.e., some behavior words are generated by the location context.
  \item[\textbf{ Assumption 6:}] The behaviors at a spot are influenced by the time, i.e., some words are generated by temporal context. 
\end{description}

\begin{figure}[!tb]
  \begin{center}
    \includegraphics[clip, width=8cm] {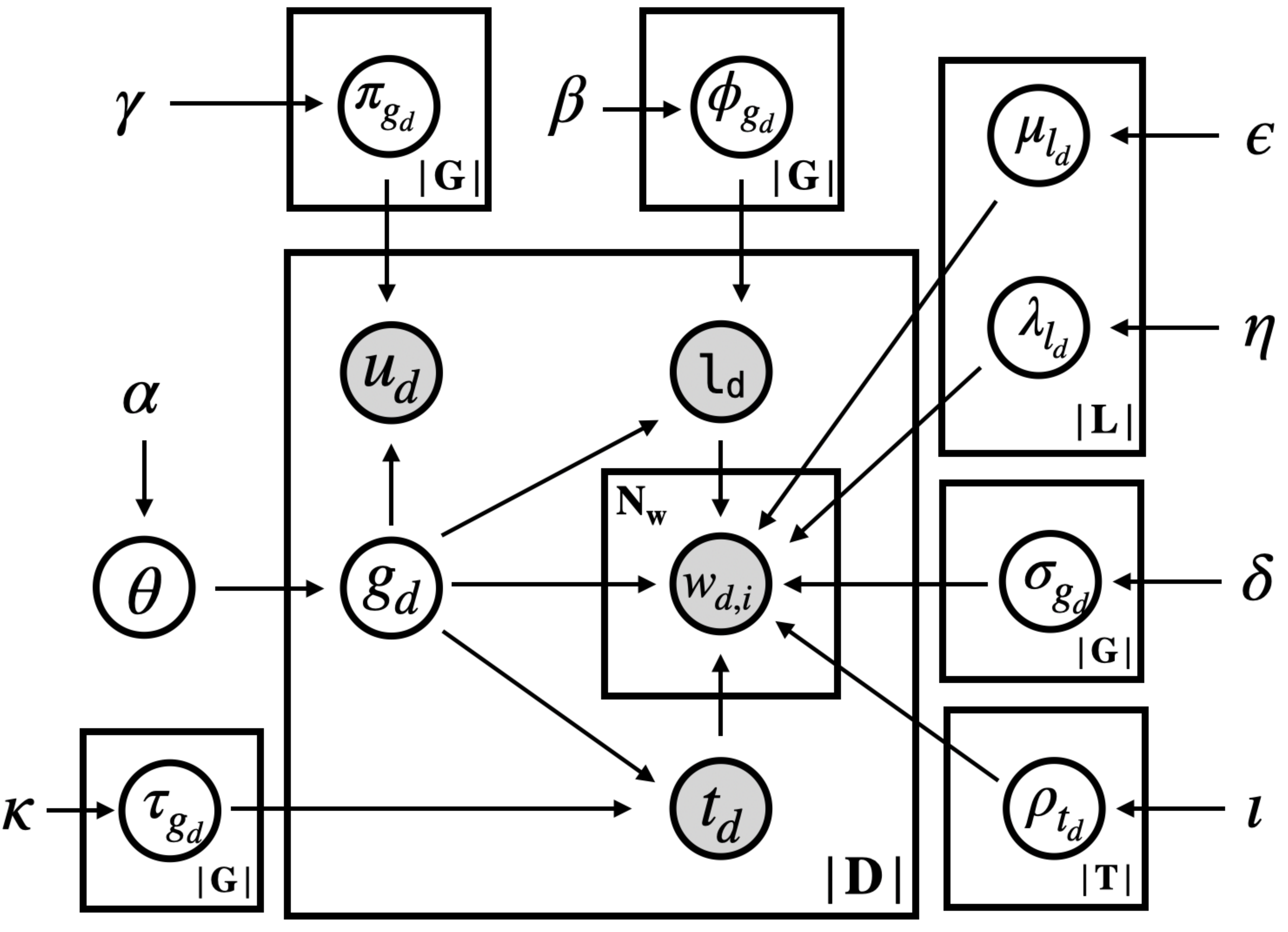}
    \caption{Graphical model of ST-UEM. Comparing with B-UEM, the word $w_{d,i}$ is influenced by spot $l_d$ and time $t_d$. }
    \label{union_model}
  \end{center}
  \end{figure}

\begin{figure}[!tb]
  \begin{center}
    \centering
    \begin{subfigure}[t]{0.47\textwidth}
      \includegraphics[clip, width=\textwidth] {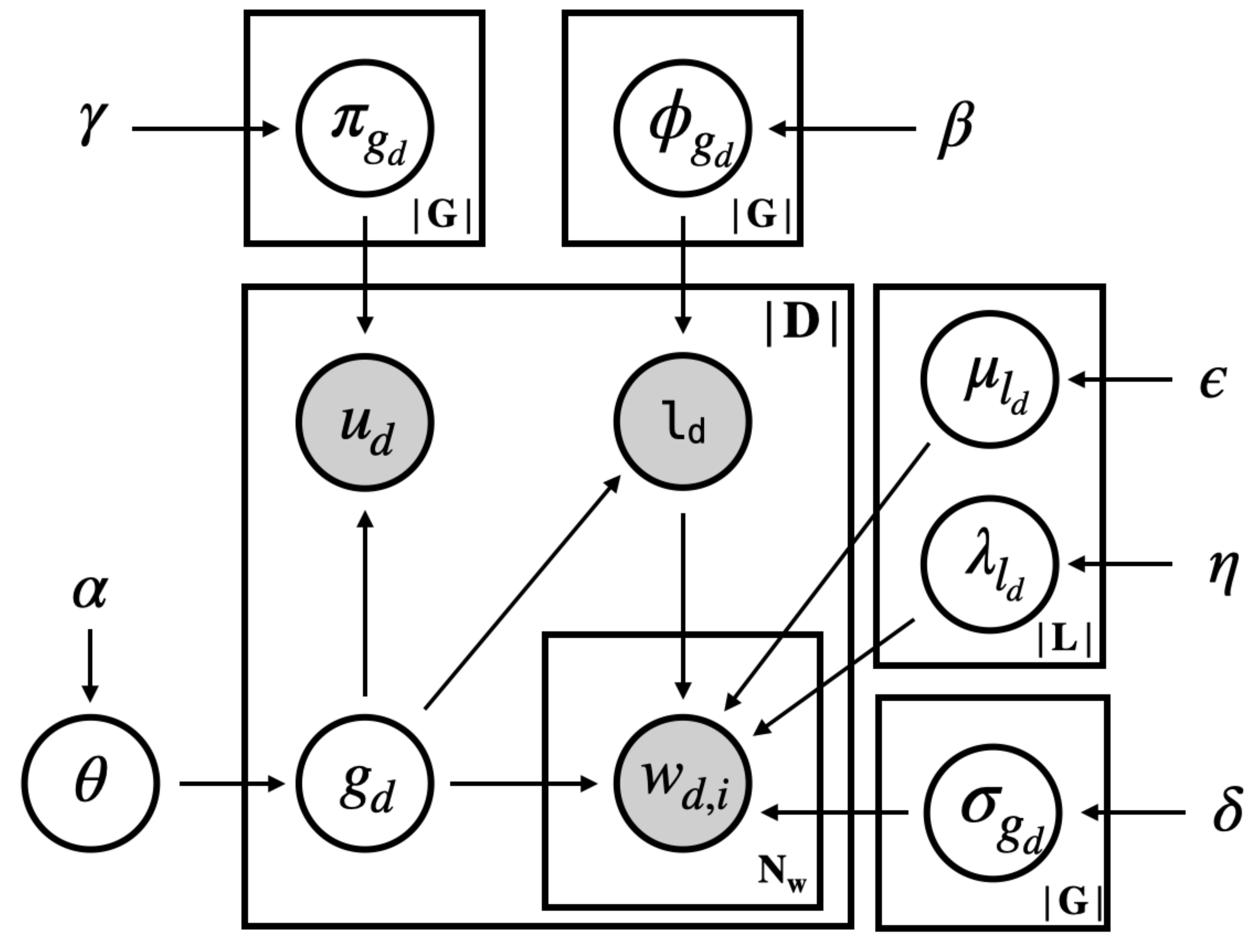}
      \caption{Graphical Model of S-UEM}
      \label{fig:S-UEM}
    \end{subfigure}
    \begin{subfigure}[t]{0.47\textwidth}
      \includegraphics[clip, width=\textwidth] {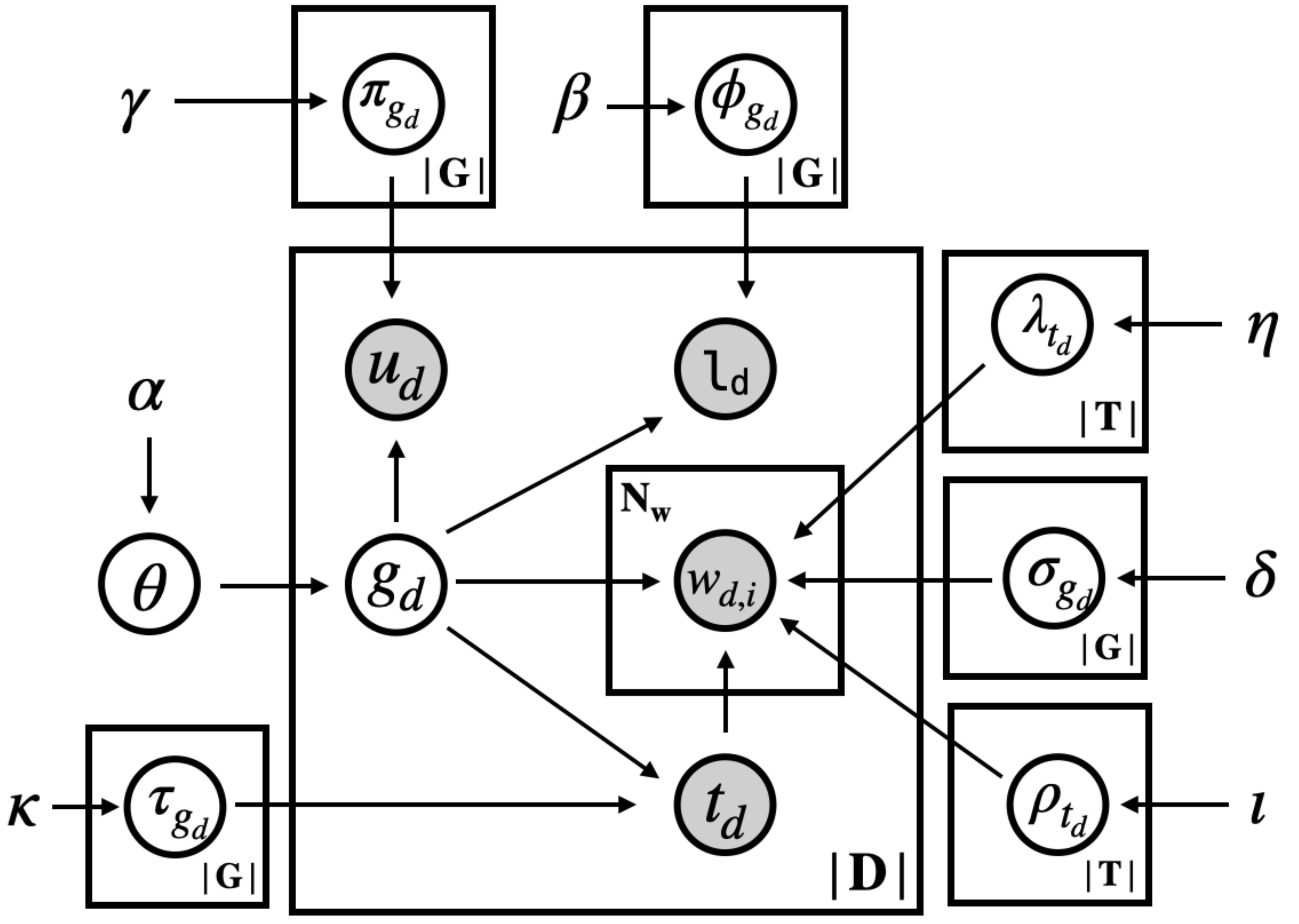}
      \caption{Graphical Model of T-UEM}
      \label{fig:T-UEM}
    \end{subfigure}
    \caption{Graphical Models of the (a) S-UEM and (b) T-UEM. Compared to the B-UEM, the S-UEM does not have a variable of time slot $t_d$, and word $w_{d,i}$ is influenced by $g_d$ and $l_d$. In the T-UEM, word $w_{d,i}$ is influenced by time slot $t_d$.}
    \label{ablation-graphical-model}
    \end{center}
  \end{figure}

With these two additional assumptions (5 and 6), we have  three variations of B-UEM: the spatio-temporal user experience model (ST-UEM), spatial user experience model (S-UEM), and temporal user experience model (T-UEM).  
The graphical model of the ST-UEM is shown in Figure \ref{union_model}. 
Graphical models of the T-UEM and S-UEM are show in Figure \ref{ablation-graphical-model}. 

\paragraph{ST-UEM}  The ST-UEM takes all the assumptions into account. Compared to the B-UEM, the behaviors and activities of tourists are spatio-temporally constrained. In other words, the generation of the word $w_d$ describing an activity is also affected by $l_d$ and $t_d$. For example, the probability of eating pizza is increased by going to an Italian restaurant at 12:00 in the afternoon or 7:00 in the evening. This is a typical example where the influence of location is strongly reflected in the behavior. Simultaneously, there are spots where the influence of time is significant. 

In this model, a variable $\lambda$ is used to model the factor, user experience, location, or time that has the greatest influence on the behavior. It is generated in a way that depends on the spot $l_d$ because $\lambda$ varies by spots. The variable $\lambda_{l_d}$ is used to generate $w_{d, i}$. Specifically, after $g_d$ and $l_d$ are generated, the variable $\lambda_{l_d}$ is generated for each location and treated as a weight to generate words describing activities.
    \begin{align}
    \lambda_{l_d} \sim& Multinomial(1, \eta_{l_d}) \\
    w_{d,i} \sim& \lambda_{l_d}^1 Categorical(\mu_{l_d})  \nonumber \\
    &+ \lambda_{l_d}^2 Categorical(\rho_{t_d}) \nonumber \\
    &+ \lambda_{l_d}^3 Categorical(\sigma_{g_d}) \\
    \mu_{l_d} \sim & Dirichlet(\epsilon)  \\
    \rho_{g_d} \sim & Dirichlet(\iota)  \\
    \sigma_{g_d} \sim & Dirichlet(\delta) 
    \end{align}
\noindent Note that $\lambda_{l_d} = (\lambda_{l_d}^1, \lambda_{l_d}^2, \lambda_{l_d}^3)$, and the first argument of the multinomial distribution is the number of times to generate.

\paragraph{S-UEM} 
The S-UEM is based on Assumptions 1, 2, 3, 4, and 5. Its graphical model is shown in Fig. \ref{fig:S-UEM}. In contrast to the B-UEM, location can determine the behavior of the tourist. Similar to the ST-UEM, $\lambda$ is introduced as a weighting factor to determine which of the two factors influences the action.
%
\paragraph{T-UEM} The T-UEM is based on Assumptions 1, 2, 3, 4, and 6. Its graphical model is shown in Fig. \ref{fig:T-UEM}. In contrast to the B-UEM, time can determine the behavior of a tourist. Compared to the ST-UEM, the influence of spot $l_d$ on word $w_d$ is removed.

\subsubsection{Experimental Setting} \label{preliminary_parameter_setting}
YFCC100M Kyoto \cite{sun2020dexa} and User Visits \cite{lim2015personalized}\cite{lim2016towards} are used. The initial hyperparameters of the Dirichlet distribution were set to 1.  $|G|$ was set to 10 as same as the similar hyperparameter of GeoMF++. The number of iterations was $5000 + 1500 * |G|$ to ensure sufficient training convergence. We split the data in two ways, {\textbf{ time-split}} and {\textbf{user-split}}. The time-split way regards tourist records of each user as a sequence and set a part of the sequence as the train data while the remaining part as the test data, to simulate the situation that we know the past records of all users and recommend them new locations. In contrast, the user-split way selects a number of users as the train group and the others as the test group to simulate the cold-start situation that we know nothing about new tourists. 

We respectively evaluate the perplexity \cite{griffiths2004finding} of locations and words describing activities. Perplexity is a measure of the accuracy of the topic model and indicates how much it can narrow down the candidates for recommendation. A higher value of perplexity indicates a larger number of choices, and a lower accuracy. 
The perplexity was calculated 10 times and averaged to eliminate the variation.

\subsubsection{Results} \label{preliminary_experiment_result}
\begin{figure}[!tb]
  \begin{center}
     \includegraphics[clip,width=0.9\textwidth] {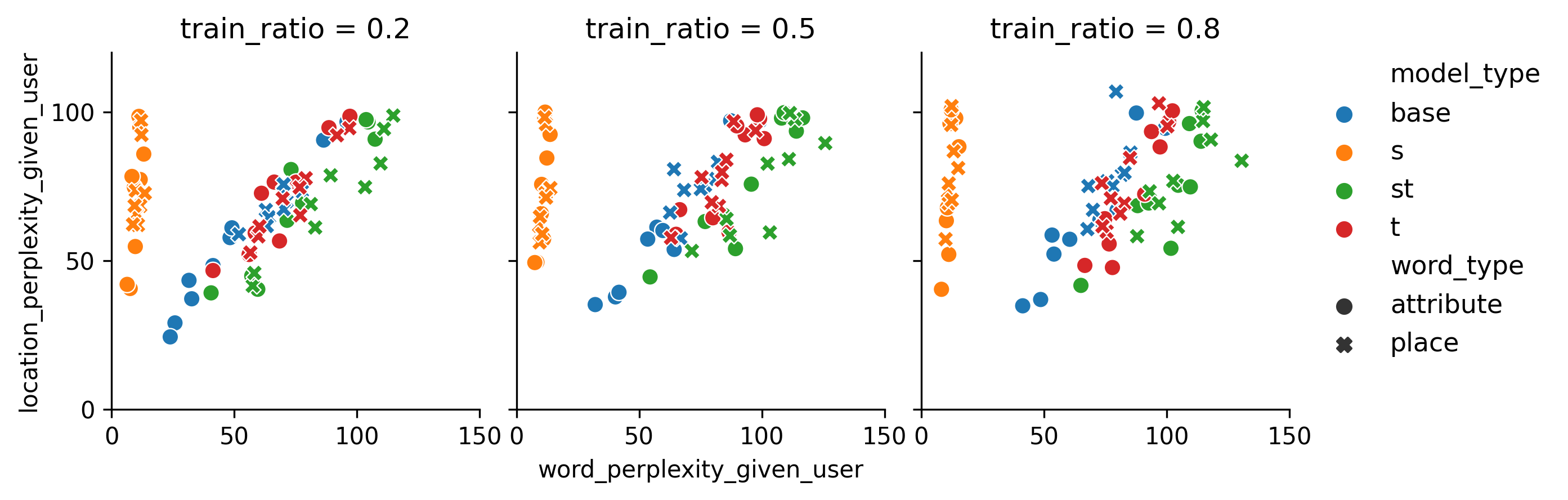}
      \caption{Perplexity of locations and words describing activities with time-split data.}
      \label{time_split_perplexity_result}
  \end{center}
  \end{figure}

\begin{figure}[!tb]
  \begin{center}
     \includegraphics[clip,width=0.9\textwidth] {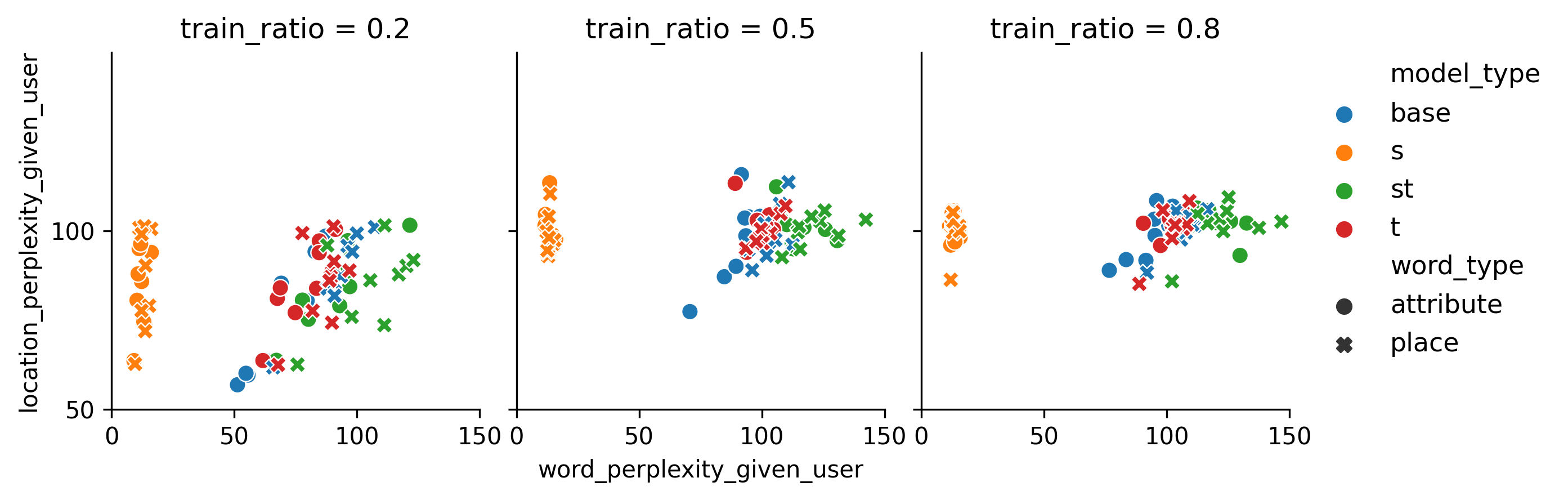}
      \caption{Perplexity of locations and words describing activities with user-split data.}
      \label{user_split_perplexity_result}
  \end{center}
  \end{figure}

Figure \ref{time_split_perplexity_result} and Figure \ref{user_split_perplexity_result} report the perplexity results of the comparative models using time-split data and user-split data, respectively. The perplexity scores show that our model performs better when using words extracted by the attribute method \cite{patterson2014sun} than the place method \cite{zhou2017places}. Hereafter, we only use the words from the attribute method. The best location perplexity performances, in order, were achieved by the B-UEM, S-UEM, ST-UEM, and T-UEM. The best word perplexity performances ,in order, were achieved by S-UEM, B-UEM, T-UEM, and ST-UEM. 

The B-UEM and S-UEM achieved the best results in location and word perplexity, respectively. This confirmed our idea that different models are needed for POI recommendation in sightseeing than for daily living. Spatio-temporal constraints (Assumptions 5 and 6) are more useful in selecting POIs in user daily mobility than in tourism. Tourists may visit sightseeing spots with a few considerations of time and locations constraints because they are in the city. However, their behaviors may be affected by the location. For example, suppose we are visiting spot A. If the ice cream there is a specialty, we may eat it even if it is not mealtime, because we may not visit spot A again.

\subsection{Evaluation of Spot Recommendations}
In this section, we discuss the comparison experiment focusing on spot recommendation against three state-of-the-art methods, GeoMF++ \cite{lian2018geomf++}, SAE-NAD \cite{Ma2018SAENAD} and LGLMF \cite{Rahmani2020LGLMF}. Precision, recall and F-value  were used as the metrics. In addition, inspired by \cite{Dacrema2019diversity}, we use the Gini index to compare the recommendation fairness. 
Gini index is a measure of statistical dispersion intended to represent the income inequality or the wealth inequality within a group\cite{gastwirth1972gini}.  We use F measure for each user as their "wealth" and apply Gini index to measure unfairness of recommendation among users. A higher value of Gini index indicates more unfairness.
\subsubsection{Parameter Setting} \label{parameter setting}
The dimension of $\alpha$ is $|G|$. The dimensions of the hyperparameters $\beta, \gamma, \delta$, and $\kappa$ are $|G| * |L|, |G| * |U|, |G| * |W|$, and $|G| * |T|$, respectively, and are affected by $G$.
The parameters have an initial value of 1.
The number of the latent user experience, $|G|$, is a hyperparameter.
Based on the model comparison evaluation and the parameter (the latent space dimensions $K$) used in GeoMF++, we set $|G|=10$. 
The number of steps depends on the size of the data and the time required for convergence. Therefore, we set the minimum and maximum number of steps to 5,000 and 150,000, respectively.

Considering that users' rating scores are usually symmetric and extreme scores are not easily given, the distribution of users' rating scores is assumed to be a 5-dimensional categorical distribution taking values $(-2, -1, 0, 1, 2)$ with probability $(F_{norm}(-1.5),  F_{norm}(-0.5) - F_{norm}(-1.5), F_{norm}(0.5) - F_{norm}(-0.5), F_{norm}(1.5) - F_{norm}(0.5), 1 - F_{norm}(1.5))$, where $F_{norm}(\cdot)$ is the cdf of the standard normal distribution.

For the GeoMF++, The default value of the latent space dimensions $K$ is 10.  
For its another hyperparameter $\alpha$, we initially performed a grid search with $\alpha={5, 10, 20, 30}$ and then let $\alpha=20$.
The maximum number of iterations was set to 15. 
For the SAE-NAD, the latent dimension of model is set to 50.
The dimension of the importance vector $d_a$ and the geographical correlation level $\gamma$ are selected by grid search, which are set to 20 and 60, respectively.
The parameters of the weighting scheme $\alpha$ and $\epsilon$ are set to 2.0 and 1e-5, respectively.
The gradien descent parameters, learning rate and regularization $\lambda$, are set to 0.001 and 0.001 respectively. 
For the LGLMF, the parameters were initialized according to the report in the corresponding paper, which means that the model uses the validation data to find the best values of the parameters and uses them in the test data. 

Similar to Section \ref{preliminary_parameter_setting} We used two data-split ways, time-split and user-split, and three train ratios, $80\%$, $50\%$ and $20\%$. The time-split way is to simulate the situation that we know the past records of all users and recommend them new locations. We perform the compare of the proposed models and baselines with time-split data in Section \ref{time_split_result}. The user-split way is to simulate the cold-start situation that we know nothing about new tourists. We perform only the compare of the proposed models with user-split data in Section \ref{user_split_result} because the existing methods cannot handle this scenario well.

\subsubsection{Experimental Results of time-split data} \label{time_split_result}
Each model was trained respectively using time-split data, and top-$k$ recommendations were made for each user for each city.
In the comparison experiment, we mainly measured the precision, recall, F measure and Gini index. 
Their values changed each time they were computed because the proposed models are probability-generating. Therefore, we ran each model and parameter setting 10 times and averaged their results.

We select two models, B-UEM and S-UEM based on the model comparison results(\ref{model_comparison}). Four recommendation methods, 
\begin{itemize}
    \item \textbf{base}: only UEM without pseudo-rating
    \item \textbf{al}:  UEM and pseudo-rating with given locations
    \item \textbf{wtol}: UEM and pseudo-rating with given words
    \item \textbf{wl}: UEM and pseudo-rating with given both words and locations
\end{itemize}
are applied to conduct recommendation tasks (Section \ref{model_application}). Considering the time split data regard tourist records as sequences and all the users' past records are known, we use the train data to estimate rating scores in al, wtol and wl methods. In al method, for any user, all the locations in the user's past records(train data) are treated as preferred locations , i.e. we suppose that users will give high scores(1 or 2) to the locations they have visited otherwise the locations would not be in their records. The wtol and wl methods are conducted similarly with replacing locations by activities and images(including both locations and activities) respectively. It is worth noting that the wtol and wl method uses the information of activities to recommend locations, which is a novel approach that few existing methods could carry out.
To evaluate the performance of methods comprehensively, we calculated the precision, recall, F measure and Gini index at top  $k$ ($k=5,10,15$) with these different train ratios.
\begin{table*}[!tb]
    \caption{Experimental Results with time-split data (Average of all cities' Precision, Recall and F-value (train ratio=0.8))}
    \label{time_split_metrics}
    \resizebox{\textwidth}{!}{
    \begin{tabular}{ c|c|c|c|c|c|c|c|c|c|c|c|c } \hline\hline
      Model & pre@5 & rec@5 & F@5 &Gini@5& pre@10 & rec@10 & F@10 &Gini@10& pre@15 & rec@15 & F@15 &Gini@15 \\  \hline
       GeoMF & 0.056 & 0.089 & 0.068 & 0.628 & 0.052 & 0.169 & 0.079 & 0.483 & 0.046 & 0.246 & 0.077 & 0.426 \\
       SAENAD & 0.104 & 0.374 & 0.162 & 0.867 & 0.083 & 0.592 & 0.145 & 0.867 & 0.069 & 0.748 & 0.126 & 0.867 \\
       LGLMF & 0.055 & 0.160 & 0.082 & 0.575 & 0.046 & 0.255 & 0.078 & 0.467 & 0.042 & 0.322 & 0.074 & 0.426 \\
       B-UEM-base & 0.135 & 0.377 & 0.198 & \textbf{0.406} & 0.112 & 0.612 & 0.188 & \textbf{0.354} & 0.095 & 0.774 & 0.170 & \textbf{0.348} \\
       B-UEM-al & \textbf{0.137} & \textbf{0.386} & \textbf{0.201} & 0.431 & \textbf{0.115} & \textbf{0.635} & \textbf{0.194} & 0.359 & \textbf{0.096} & \textbf{0.788} & \textbf{0.172} & 0.351 \\
       B-UEM-wtol & 0.122 & 0.345 & 0.180 & 0.430 & 0.103 & 0.572 & 0.175 & 0.368 & 0.091 & 0.742 & 0.161 & 0.356 \\
       B-UEM-wl & 0.129 & 0.366 & 0.190 & 0.421 & 0.108 & 0.599 & 0.183 & 0.364 & 0.093 & 0.763 & 0.166 & 0.352 \\
       S-UEM-base & 0.136 & 0.381 & 0.199 & \textbf{0.407} & 0.111 & 0.612 & 0.188 & \textbf{0.354} & 0.095 & 0.771 & 0.169 & \textbf{0.347} \\
       S-UEM-al & \textbf{0.138} & \textbf{0.391} & \textbf{0.204} & 0.432 & \textbf{0.114} & \textbf{0.632} & \textbf{0.193} & 0.359 & \textbf{0.096} & \textbf{0.782} & \textbf{0.171} & 0.352 \\
       S-UEM-wtol & 0.130 & 0.366 & 0.191 & 0.433 & 0.109 & 0.597 & 0.184 & 0.366 & 0.094 & 0.759 & 0.166 & 0.356 \\
       S-UEM-wl & 0.136 & 0.385 & 0.200 & 0.427 & 0.112 & 0.618 & 0.189 & 0.362 & 0.095 & 0.773 & 0.169 & 0.354 \\ \hline\hline
    \end{tabular}
    }
\end{table*}
As shown in Table \ref{time_split_metrics}, our methods outperformed GeoMF++, SAE-NAD and LGLMF in precision, reccall ,F measure and Gini index, which means that our methods performed better POI recommendation in average and achieve more fairness among users. The B-UEM-al method and S-UEM-al are superior to the others in precision, recall and F measure, while B-UEM-base and S-UEM-base achieved fairer results denoted by Gini index.
 
\begin{figure}[!htb]
    \begin{center}
     \includegraphics[clip,width=0.45\textwidth] {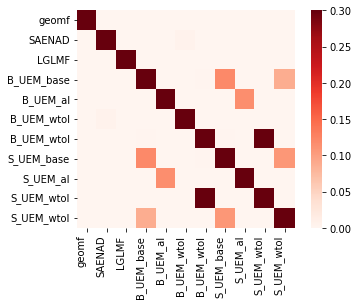}
      \caption{Wilcoxon signed-rank test of Average F-measure with time-split data}
      \label{time_split_significance}
   \end{center}
  \end{figure}

We also test the significance of the difference of F measure between our models and the baselines, the results are shown in Fig. \ref{time_split_significance}. The difference between any proposed method and any baseline is significant, which proved that our UEM based methods achieved significant improvements compared to the baselines.  
Different from the results shown in model comparison (Fig. \ref{time_split_perplexity_result}, Fig. \ref{user_split_perplexity_result}), Table \ref{time_split_metrics} and Fig. \ref{time_split_significance} reveals that there is no significant difference between the B-UEM and S-UEM.
One of the considerable reasons is that although the S-UEM will recommend the best place for a certain experience by considering the sptaial context, a tourist may experience something in a convenient place and the B-UEM could handle this situation well.
The compare of al, wtol, wl methods also illustrates this situation. In contrast to al method, the wtol method uses start-up information of words describing activities and the wl method uses both words describing activities and locations. However, the al method outperformed wtol and wl methods, which indicates that the relationship between activities and locations in tourism recommendation is not as close as daily life.
In addition, the B-UEM is preferred to the S-UEM because it is simpler and easier to train and achieved better performance in the predication of user behaviors.
\begin{figure*}[!htb]
\begin{center}
    \begin{subfigure}[!tb]{0.33\textwidth}
      \includegraphics[clip, width=\textwidth] {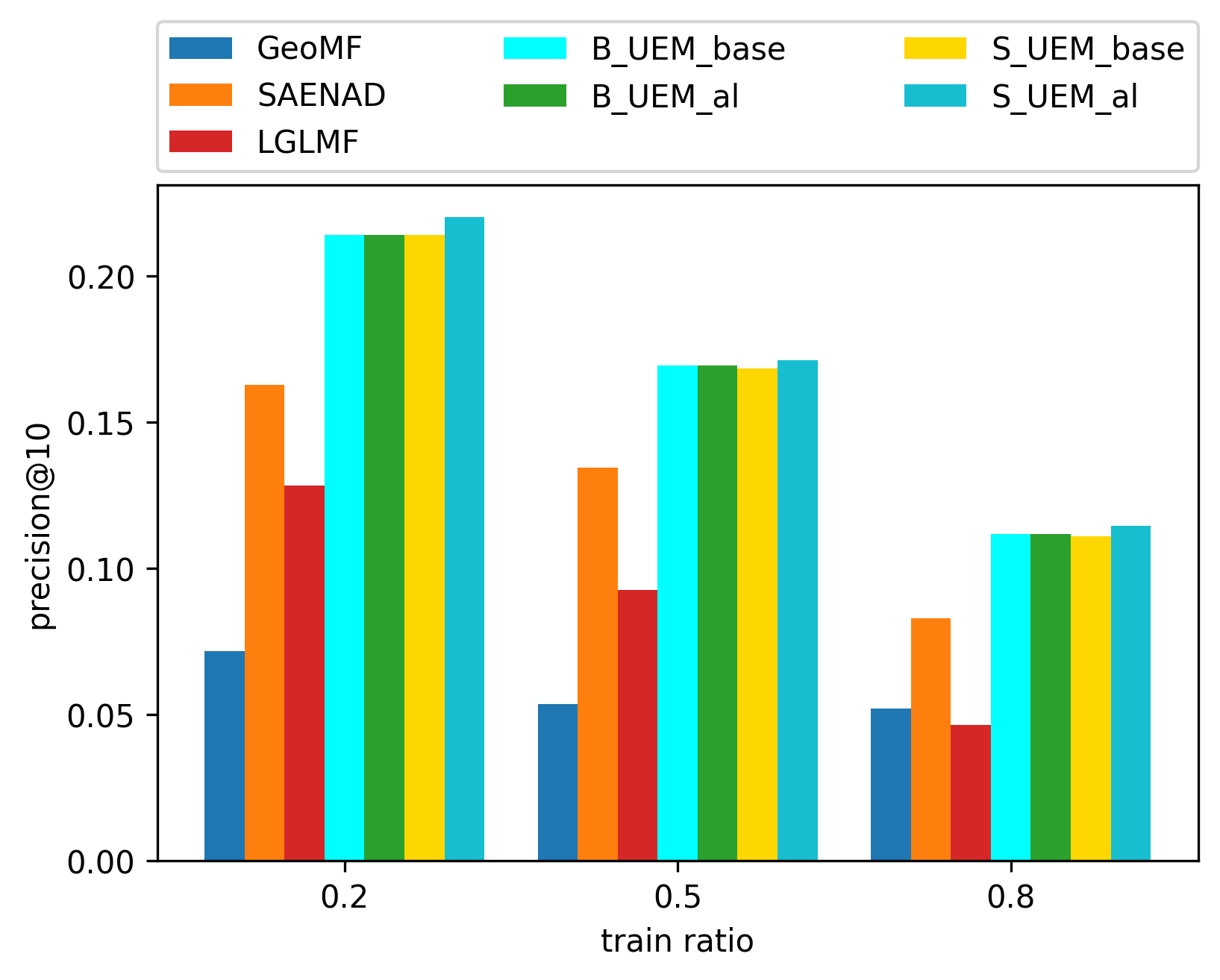}
      \caption{}
    \end{subfigure}
    \begin{subfigure}[!tb]{0.33\textwidth}
      \includegraphics[clip, width=\textwidth] {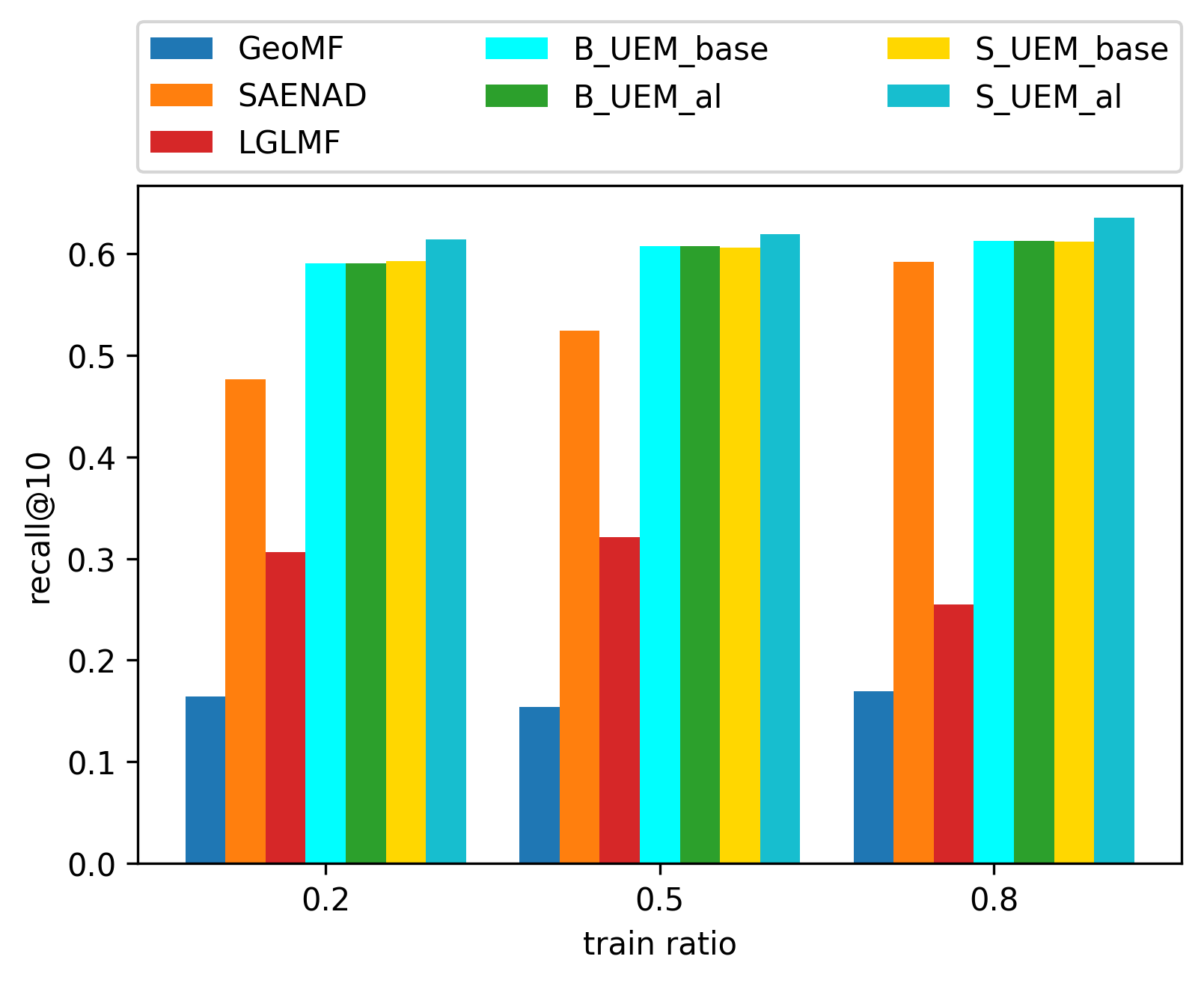}
      \caption{}
    \end{subfigure}
    \begin{subfigure}[!tb]{0.33\textwidth}
      \includegraphics[clip, width=\textwidth] {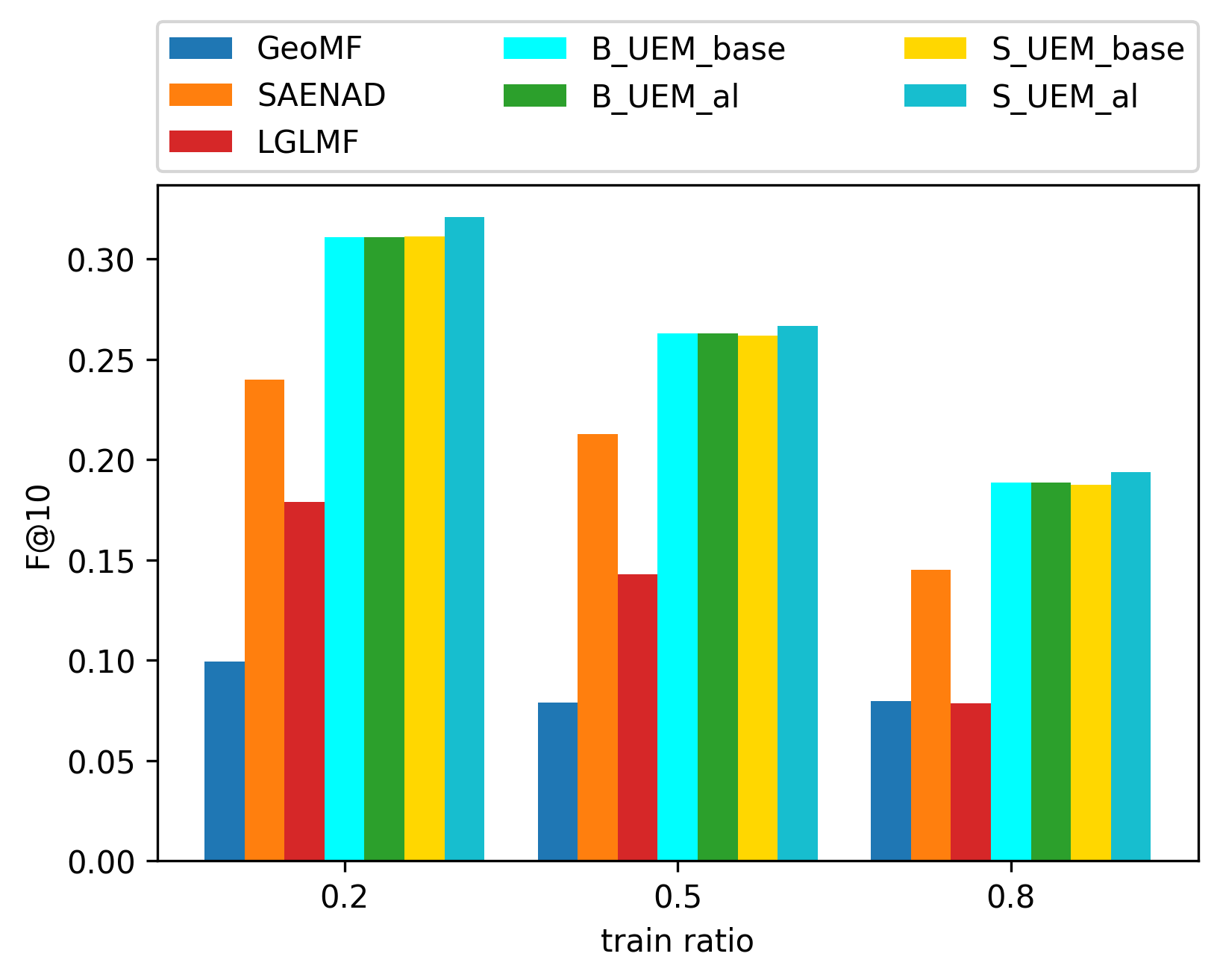}
      \caption{}
    \end{subfigure}
    \caption{Average of all cities' (a)Precision@10, (b) Recall@10 and (c)F@10 with different training ratios of time-split data. }
    \label{train_ratio_time_split_figure}
\end{center}
\end{figure*}

For further validation, we experimented with different ratios of training data among total data, and the average of all cities' precision, recall and F measure of the top 10 recommendations are shown in Figure \ref{train_ratio_time_split_figure}. The proposed methods achieved similar results with any train ratio, which indicates that our model is robust even when the training data is sparse. 
On the other hand, the UEM based methods outperformed the baselines in terms of both precision, recall and F measure for any portion of the training data. This reveals that our model is more robust and can handle sparse data.

\subsubsection{Experimental Results of user-split data} \label{user_split_result}
Similar to experiment using time-split data, each model was trained using user-split data and top-$k$ recommendations were made for each user for each city. We mainly measured the precision, recall, F measure and Gini index. 
The user-split data simulate the situation that we have learned behavior patterns from some old users and know nothing about new users and no existing methods can handle this scenario well. Hence, we only performed the compare of the proposed models in this section. We compared B-UEM and S-UEM with three recommendation methods, \textbf{al}(UEM and pseudo-rating with given locations), \textbf{wtol}(UEM and pseudo-rating with given words), \textbf{wl}(UEM and pseudo-rating with given both words and locations). We ran each model and parameter setting 10 times and averaged their results. 

\begin{figure}[!htb]
   \begin{center}
     \includegraphics[clip,width=0.45\textwidth] {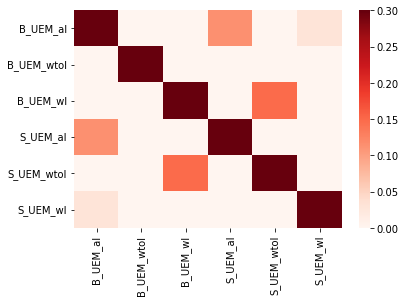}
      \caption{User-split data}
      \label{user_split_significance}
  \end{center}
  \end{figure}

For any user in test data, we pick five locations or activities, one location or activity that the user has experienced and four locations or activities that the user has not experienced, as start-up information because we could not learn anything about the user from training data. As previously mentioned, we use the start-up information to simulate that five locations or activities are provided to the user, and the user gives rating scores according to the actual tourism records. The location or activity that the user has experienced is treated as preferred location or activity, i.e. score $1$ or $2$, and the location or activity that the user has not experienced are treated as disfavored location or activity, i.e. score $-2, -1$ or $0$. 
\begin{table*}[!htb]
    \caption{Experimental Results with user-split data (Average of all cities' Precision, Recall and F-value (train ratio=0.8))}
    \label{user_split_metrics}
    \resizebox{\textwidth}{!}{
    \begin{tabular}{ c|c|c|c|c|c|c|c|c|c|c|c|c } \hline\hline
      Model & pre@5 & rec@5 & F@5 &Gini@5& pre@10 & rec@10 & F@10 &Gini@10& pre@15 & rec@15 & F@15 &Gini@15 \\  \hline
      B-UEM-al & \textbf{0.300} & \textbf{0.355} & \textbf{0.321} & 0.475 & \textbf{0.258} & \textbf{0.592} & \textbf{0.357} & \textbf{0.434} & \textbf{0.224} & 0.764 & \textbf{0.343} & 0.434 \\
      B-UEM-wtol & 0.279 & 0.326 & 0.297 & 0.492 & 0.248 & 0.567 & 0.342 & 0.445 & 0.219 & 0.751 & 0.336 & 0.437 \\
      B-UEM-wl & 0.295 & 0.348 & 0.315 & \textbf{0.472} & 0.254 & 0.581 & 0.350 & \textbf{0.433} & 0.221 & 0.756 & 0.339 & \textbf{0.433} \\
      S-UEM-al & \textbf{0.303} & \textbf{0.360} & \textbf{0.326} & 0.475 & \textbf{0.259} & \textbf{0.596} & \textbf{0.358} & 0.435 & \textbf{0.224} & \textbf{0.769} & \textbf{0.344} & \textbf{0.433} \\
      S-UEM-wtol & 0.285 & 0.332 & 0.302 & 0.500 & 0.254 & 0.580 & 0.350 & 0.447 & 0.222 & 0.759 & 0.341 & 0.437 \\
      S-UEM-wl & 0.300 & 0.358 & 0.323 & \textbf{0.472} & 0.256 & 0.589 & 0.354 & \textbf{0.434} & 0.223 & \textbf{0.765} & 0.342 & \textbf{0.432} \\ \hline\hline
    \end{tabular}
    }
\end{table*}
As shown in Table \ref{user_split_metrics}, the B-UEM-al and S-UEM-al outperform the other methods in average precision, recall and F measure, while B-UEM-wl and S-UEM-wl perform better than the other methods in the Gini index. Fig. \ref{user_split_significance} reveals that there is no significant difference between the B-UEM-al and S-UEM-al, while the differences between B-UEM-al or S-UEM-al and other methods are significant. 

\begin{figure*}[!htb]
  \begin{center}
    \begin{subfigure}[t]{0.33\textwidth}
      \includegraphics[clip, width=\textwidth] {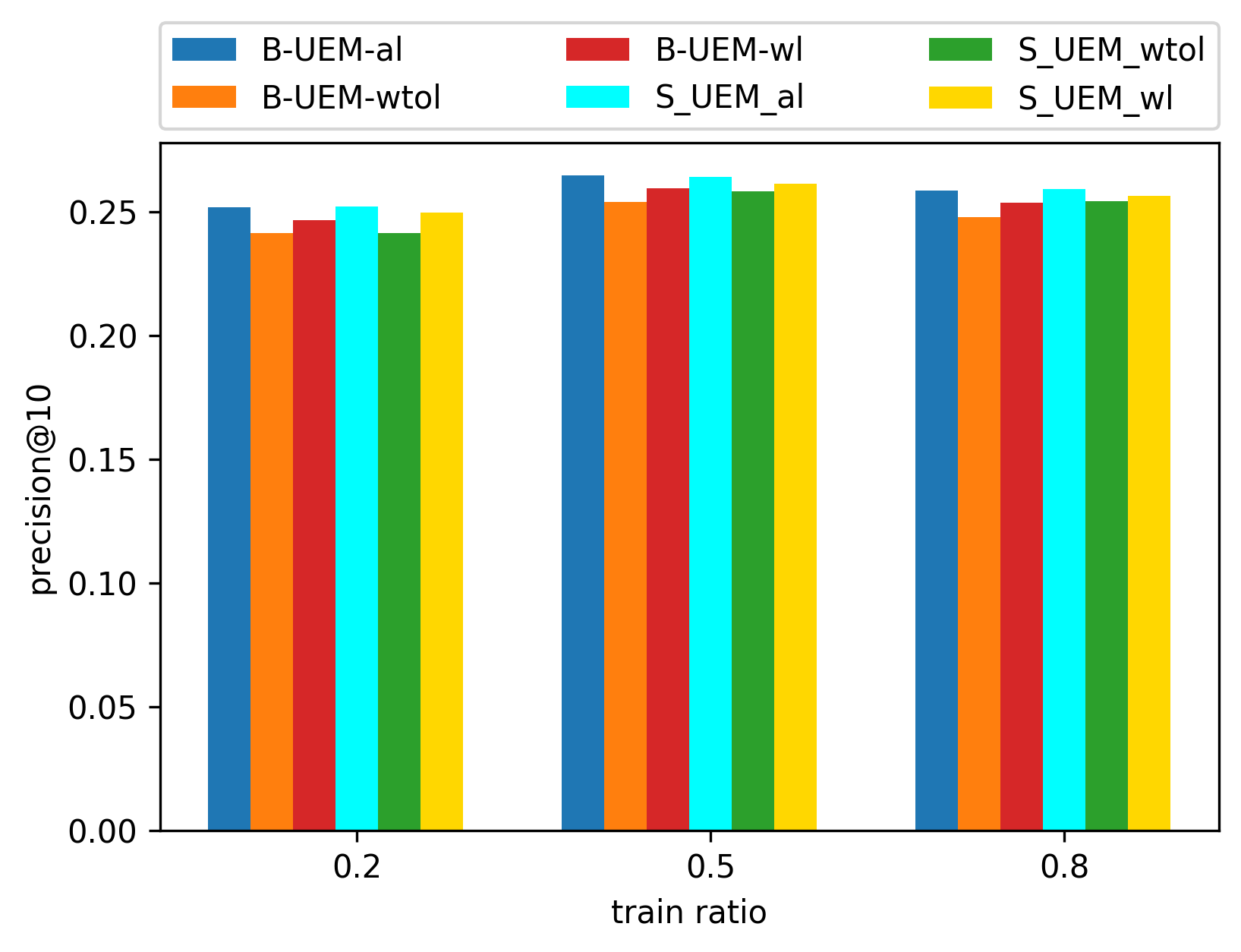}
      \caption{}
    \end{subfigure}
    \begin{subfigure}[t]{0.33\textwidth}
      \includegraphics[clip, width=\textwidth] {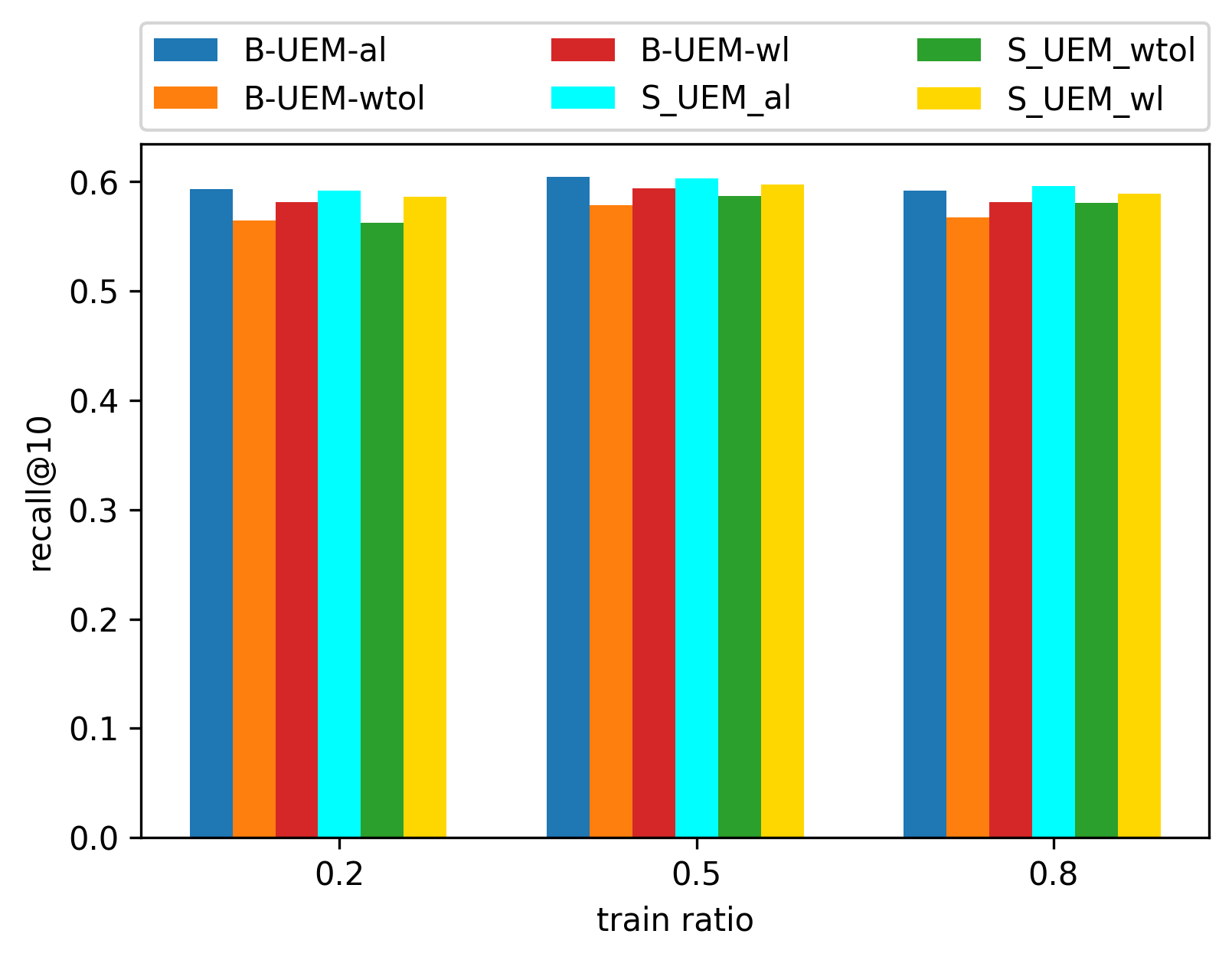}
      \caption{}
    \end{subfigure}
    \begin{subfigure}[t]{0.33\textwidth}
      \includegraphics[clip, width=\textwidth] {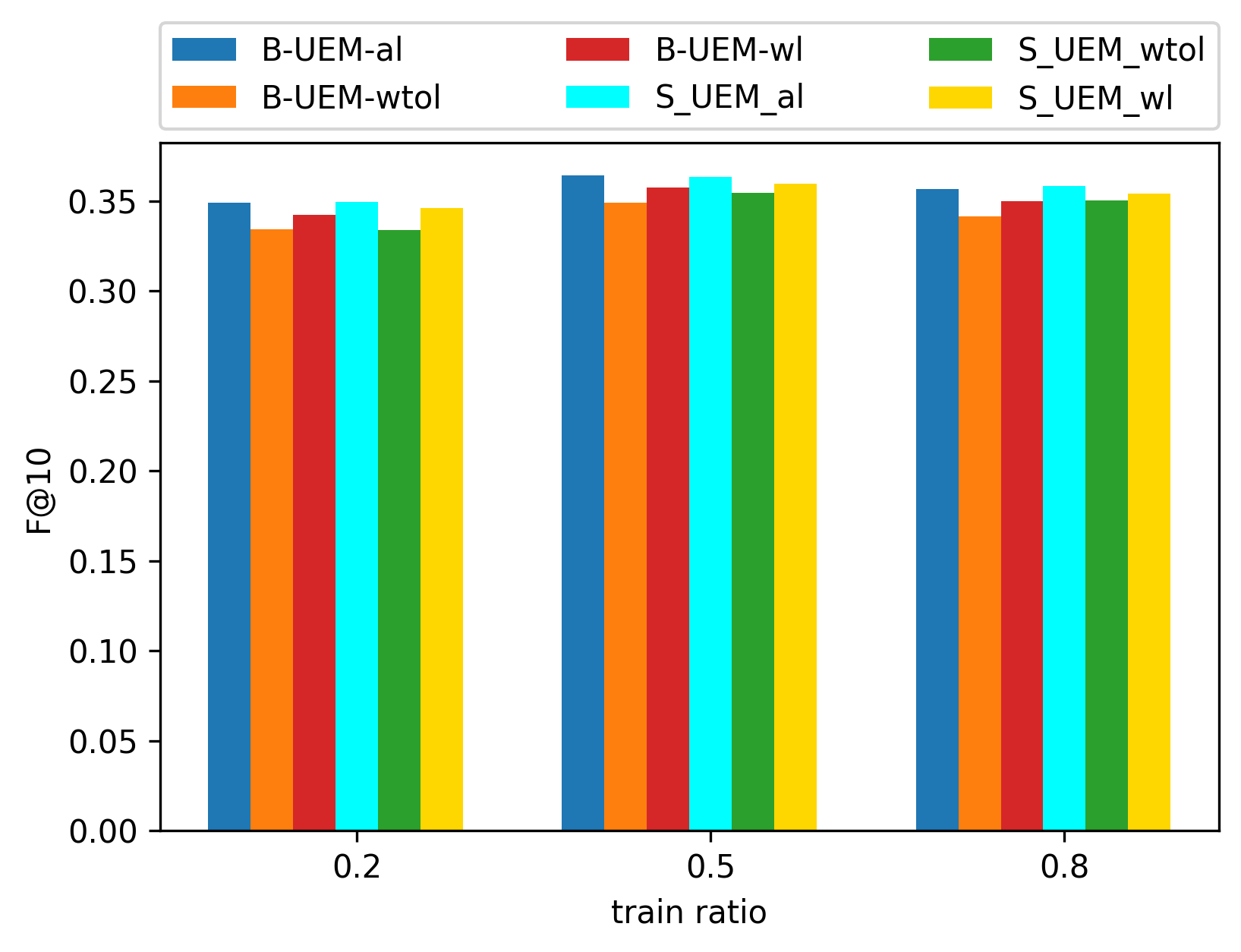}
      \caption{}
    \end{subfigure}
\caption{Average of all cities' (a)Precision@10, (b) Recall@10 and (c)F@10 with different training ratios of user-split data. }
    \label{train_ratio_user_split_figure}
  \end{center}
  \end{figure*}

Fig. \ref{train_ratio_user_split_figure} shows that the proposed methods achieved similar results with differenet train ratios, which indicates that our model is robust and can handle sparse data.
We also test the influence of the assumption about users' rating distribution. To test the assumption about symmetry, we used another 5-dimensional categorical distribution taking values $(-2, -1, 0, 1, 2)$ with probability $(F_{exp}(-1.5), F_{exp}(-0.5) - F_{exp}(-1.5), F_{exp}(0.5) - F_{exp}(-0.5), F_{exp}(1.5) - F_{exp}(0.5), 1 - F_{exp}(1.5))$, where $F_{exp}(\cdot)$ is the cdf of the exponential distribution with shifting location $-2.5$ and parameter $1/\sigma$. To test the probabilities settings, we let the standard deviation $\sigma$ varies from $0.01$ to $2$. 
\begin{figure}[!htb]
\begin{center}
     \includegraphics[clip,width=0.5\textwidth] {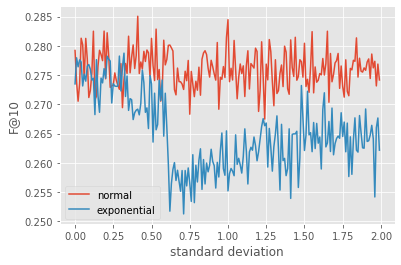}
      \caption{The F measure (k=10) using user-split data with different distribution assumption.}
      \label{user_split_distribution}
\end{center}
\end{figure}

As shown in Fig. \ref{user_split_distribution}
, the average $F@10$ using symmetric distribution outperformed that using asymmetric distribution, which suggests that the assumption of symmetric is practical. On the other hand, the influence of standard deviation seems to be affected by the assumption of symmetric. There are no significant changes in symmetric case while a sharp decrease can be confirmed in asymmetric case, which suggests that setting smaller probabilities for extreme rates lead to better recommendation. 

\section{Conclusions} \label{conclusion}
This study defined a new class of POI recommendations in tourism. We proposed a novel user-experience model to analyze tourist behavior with extremely sparse data. We also propose spots recommendation methods by revealing the activities tourists. In addition, a novel pseudo rating mechanism is proposed to handle the cold-start scenario, in which tourists are new to city and have no historical data. Extensive experiments on two real-data set have been carried out. The experimental results validate that our assumption that user behaviors in sightseeing differ from those in daily life and thus, different models to support tourism are needed. Our method outperformed the state-of-the-art baselines in both effectiveness and fairness and could make satisfactory recommendation to new tourists, even with extremely sparse data.  

\section*{Acknoledgements}
This work was supported by MIC SCOPE (201607008).



\begin{thebibliography}{10}

\bibitem{zhuang2017sns}
Chenyi Zhuang, Qiang Ma, and Masatoshi Yoshikawa.
\newblock Sns user classification and its application to obscure poi discovery.
\newblock {\em Multimedia Tools and Applications}, 76(4):5461--5487, 2017.

\bibitem{DBLP:journals/mta/GeZM19}
Min Ge, Chenyi Zhuang, and Qiang Ma.
\newblock Robust visual object clustering and its application to sightseeing
  spot assessment.
\newblock {\em Multim. Tools Appl.}, 78(12):17135--17164, 2019.

\bibitem{DBLP:journals/ijbdi/ShenGZM18}
Yizhu Shen, Min Ge, Chenyi Zhuang, and Qiang Ma.
\newblock Sightseeing value estimation by analysing geosocial images.
\newblock {\em Int. J. Big Data Intell.}, 5(1/2):31--48, 2018.

\bibitem{lim2015personalized}
Kwan~Hui Lim, Jeffrey Chan, Christopher Leckie, and Shanika Karunasekera.
\newblock Personalized tour recommendation based on user interests and points
  of interest visit durations.
\newblock In {\em Twenty-Fourth International Joint Conference on Artificial
  Intelligence}, 2015.

\bibitem{zhou2017places}
Bolei Zhou, Agata Lapedriza, Aditya Khosla, Aude Oliva, and Antonio Torralba.
\newblock Places: A 10 million image database for scene recognition.
\newblock {\em IEEE Transactions on Pattern Analysis and Machine Intelligence},
  2017.

\bibitem{10.1007/s11042-010-0623-y}
Jiebo Luo, Dhiraj Joshi, Jie Yu, and Andrew Gallagher.
\newblock Geotagging in multimedia and computer vision--a survey.
\newblock {\em Multimedia Tools Appl.}, 51(1):187–211, January 2011.

\bibitem{DBLP:journals/mta/ZhengZC11}
Yan{-}Tao Zheng, Zheng{-}Jun Zha, and Tat{-}Seng Chua.
\newblock Research and applications on georeferenced multimedia: a survey.
\newblock {\em Multim. Tools Appl.}, 51(1):77--98, 2011.

\bibitem{islam2020survey}
Md.~Ashraful Islam, Mir~Mahathir Mohammad, Sarkar Snigdha~Sarathi Das, and
  Mohammed~Eunus Ali.
\newblock A survey on deep learning based point-of-interest (poi)
  recommendations, 2020.

\bibitem{yuan2013and}
Quan Yuan, Gao Cong, Zongyang Ma, Aixin Sun, and Nadia~Magnenat Thalmann.
\newblock Who, where, when and what: discover spatio-temporal topics for
  twitter users.
\newblock In {\em Proceedings of the 19th ACM SIGKDD international conference
  on Knowledge discovery and data mining}, pages 605--613. ACM, 2013.

\bibitem{yuan2015and}
Quan Yuan, Gao Cong, Kaiqi Zhao, Zongyang Ma, and Aixin Sun.
\newblock Who, where, when, and what: A nonparametric bayesian approach to
  context-aware recommendation and search for twitter users.
\newblock {\em ACM Transactions on Information Systems (TOIS)}, 33(1):2, 2015.

\bibitem{ference2013location}
Gregory Ference, Mao Ye, and Wang-Chien Lee.
\newblock Location recommendation for out-of-town users in location-based
  social networks.
\newblock In {\em Proceedings of the 22nd ACM international conference on
  Information \& Knowledge Management}, pages 721--726, 2013.

\bibitem{lian2014geomf}
Defu Lian, Cong Zhao, Xing Xie, Guangzhong Sun, Enhong Chen, and Yong Rui.
\newblock Geomf: joint geographical modeling and matrix factorization for
  point-of-interest recommendation.
\newblock In {\em Proceedings of the 20th ACM SIGKDD international conference
  on Knowledge discovery and data mining}, pages 831--840, 2014.

\bibitem{lian2018geomf++}
Defu Lian, Kai Zheng, Yong Ge, Longbing Cao, Enhong Chen, and Xing Xie.
\newblock Geomf++ scalable location recommendation via joint geographical
  modeling and matrix factorization.
\newblock {\em ACM Transactions on Information Systems (TOIS)}, 36(3):1--29,
  2018.

\bibitem{DBLP:journals/www/XuFCLW20}
Shuai Xu, Xiaoming Fu, Jiuxin Cao, Bo~Liu, and Zhixiao Wang.
\newblock Survey on user location prediction based on geo-social networking
  data.
\newblock {\em World Wide Web}, 23(3):1621--1664, 2020.

\bibitem{DBLP:journals/kais/LimCKL19}
Kwan~Hui Lim, Jeffrey Chan, Shanika Karunasekera, and Christopher Leckie.
\newblock Tour recommendation and trip planning using location-based social
  media: a survey.
\newblock {\em Knowl. Inf. Syst.}, 60(3):1247--1275, 2019.

\bibitem{Rahmani2020LGLMF}
Hossein~A. Rahmani, Mohammad Aliannejadi, Sajad Ahmadian, Mitra Baratchi,
  Mohsen Afsharchi, and Fabio Crestani.
\newblock Lglmf: Local geographical based logistic matrix factorization model
  for poi recommendation.
\newblock In Fu~Lee Wang, Haoran Xie, Wai Lam, Aixin Sun, Lun-Wei Ku, Tianyong
  Hao, Wei Chen, Tak-Lam Wong, and Xiaohui Tao, editors, {\em Information
  Retrieval Technology}, pages 66--78, Cham, 2020. Springer International
  Publishing.

\bibitem{Ma2018SAENAD}
Chen Ma, Yingxue Zhang, Qinglong Wang, and Xue Liu.
\newblock Point-of-interest recommendation: Exploiting self-attentive
  autoencoders with neighbor-aware influence.
\newblock In {\em {CIKM}}, pages 697--706. {ACM}, 2018.

\bibitem{lim2016towards}
Kwan~Hui Lim, Jeffrey Chan, Christopher Leckie, and Shanika Karunasekera.
\newblock Towards next generation touring: Personalized group tours.
\newblock In {\em Proceedings of the International Conference on Automated
  Planning and Scheduling}, volume~26, 2016.

\bibitem{yin2016joint}
Hongzhi Yin, Bin Cui, Xiaofang Zhou, Weiqing Wang, Zi~Huang, and Shazia Sadiq.
\newblock Joint modeling of user check-in behaviors for real-time
  point-of-interest recommendation.
\newblock {\em ACM Transactions on Information Systems (TOIS)}, 35(2):1--44,
  2016.

\bibitem{yin2013lcars}
Hongzhi Yin, Yizhou Sun, Bin Cui, Zhiting Hu, and Ling Chen.
\newblock Lcars: a location-content-aware recommender system.
\newblock In {\em Proceedings of the 19th ACM SIGKDD international conference
  on Knowledge discovery and data mining}, pages 221--229, 2013.

\bibitem{yin2015joint}
Hongzhi Yin, Xiaofang Zhou, Yingxia Shao, Hao Wang, and Shazia Sadiq.
\newblock Joint modeling of user check-in behaviors for point-of-interest
  recommendation.
\newblock In {\em Proceedings of the 24th ACM International on Conference on
  Information and Knowledge Management}, pages 1631--1640, 2015.

\bibitem{liu2013personalized}
Xin Liu, Yong Liu, Karl Aberer, and Chunyan Miao.
\newblock Personalized point-of-interest recommendation by mining users'
  preference transition.
\newblock In {\em Proceedings of the 22nd ACM international conference on
  Information \& Knowledge Management}, pages 733--738, 2013.

\bibitem{sun2020dexa}
Junjie Sun, Tomoki Kinoue, and Qiang Ma.
\newblock A city adaptive clustering framework for discovering pois with
  different granularities.
\newblock In Sven Hartmann, Josef K{\"u}ng, Gabriele Kotsis, A.~Min Tjoa, and
  Ismail Khalil, editors, {\em Database and Expert Systems Applications}, pages
  425--434, Cham, 2020. Springer International Publishing.

\bibitem{abdollahpouri2019unfairness}
Himan Abdollahpouri, Masoud Mansoury, Robin Burke, and Bamshad Mobasher.
\newblock The unfairness of popularity bias in recommendation.
\newblock {\em arXiv preprint arXiv:1907.13286}, 2019.

\bibitem{ashkan2014diversified}
Azin Ashkan, Branislav Kveton, Shlomo Berkovsky, and Zheng Wen.
\newblock Diversified utility maximization for recommendations.
\newblock In {\em RecSys Posters}. Citeseer, 2014.

\bibitem{steck2018calibrated}
Harald Steck.
\newblock Calibrated recommendations.
\newblock In {\em Proceedings of the 12th ACM conference on recommender
  systems}, pages 154--162, 2018.

\bibitem{song2010limits}
Chaoming Song, Zehui Qu, Nicholas Blumm, and Albert-L{\'a}szl{\'o}
  Barab{\'a}si.
\newblock Limits of predictability in human mobility.
\newblock {\em Science}, 327(5968):1018--1021, 2010.

\bibitem{gao2013exploring}
Huiji Gao, Jiliang Tang, Xia Hu, and Huan Liu.
\newblock Exploring temporal effects for location recommendation on
  location-based social networks.
\newblock In {\em Proceedings of the 7th ACM conference on Recommender
  systems}, pages 93--100, 2013.

\bibitem{gao2013modeling}
Huiji Gao, Jiliang Tang, Xia Hu, and Huan Liu.
\newblock Modeling temporal effects of human mobile behavior on location-based
  social networks.
\newblock In {\em Proceedings of the 22nd ACM international conference on
  Information \& Knowledge Management}, pages 1673--1678, 2013.

\bibitem{wang2015geo}
Weiqing Wang, Hongzhi Yin, Ling Chen, Yizhou Sun, Shazia Sadiq, and Xiaofang
  Zhou.
\newblock Geo-sage: A geographical sparse additive generative model for spatial
  item recommendation.
\newblock In {\em Proceedings of the 21th ACM SIGKDD International Conference
  on Knowledge Discovery and Data Mining}, pages 1255--1264. ACM, 2015.

\bibitem{liu2017experimental}
Yiding Liu, Tuan-Anh~Nguyen Pham, Gao Cong, and Quan Yuan.
\newblock An experimental evaluation of point-of-interest recommendation in
  location-based social networks.
\newblock {\em Proceedings of the VLDB Endowment}, 10(10):1010--1021, 2017.

\bibitem{yin2016adapting}
Hongzhi Yin, Xiaofang Zhou, Bin Cui, Hao Wang, Kai Zheng, and Quoc Viet~Hung
  Nguyen.
\newblock Adapting to user interest drift for poi recommendation.
\newblock {\em IEEE Transactions on Knowledge and Data Engineering},
  28(10):2566--2581, 2016.

\bibitem{aliannejadi2018personalized}
Mohammad Aliannejadi and Fabio Crestani.
\newblock Personalized context-aware point of interest recommendation.
\newblock {\em ACM Transactions on Information Systems (TOIS)}, 36(4):1--28,
  2018.

\bibitem{rahimi2019behavior}
Seyyed~Mohammadreza Rahimi, Behrouz Far, and Xin Wang.
\newblock Behavior-based location recommendation on location-based social
  networks.
\newblock {\em GeoInformatica}, pages 1--28, 2019.

\bibitem{wang2007mining}
Chong Wang, Jinggang Wang, Xing Xie, and Wei-Ying Ma.
\newblock Mining geographic knowledge using location aware topic model.
\newblock In {\em Proceedings of the 4th ACM workshop on Geographical
  information retrieval}, pages 65--70, 2007.

\bibitem{hong2012discovering}
Liangjie Hong, Amr Ahmed, Siva Gurumurthy, Alexander~J Smola, and Kostas
  Tsioutsiouliklis.
\newblock Discovering geographical topics in the twitter stream.
\newblock In {\em Proceedings of the 21st international conference on World
  Wide Web}, pages 769--778, 2012.

\bibitem{wang2018tpm}
Weiqing Wang, Hongzhi Yin, Xingzhong Du, Quoc Viet~Hung Nguyen, and Xiaofang
  Zhou.
\newblock Tpm: A temporal personalized model for spatial item recommendation.
\newblock {\em ACM Transactions on Intelligent Systems and Technology (TIST)},
  9(6):1--25, 2018.

\bibitem{yin2016discovering}
Hongzhi Yin, Zhiting Hu, Xiaofang Zhou, Hao Wang, Kai Zheng, Quoc Viet~Hung
  Nguyen, and Shazia Sadiq.
\newblock Discovering interpretable geo-social communities for user behavior
  prediction.
\newblock In {\em 2016 IEEE 32nd International Conference on Data Engineering
  (ICDE)}, pages 942--953. IEEE, 2016.

\bibitem{thomee2016yfcc100m}
Bart Thomee, David~A Shamma, Gerald Friedland, Benjamin Elizalde, Karl Ni,
  Douglas Poland, Damian Borth, and Li-Jia Li.
\newblock Yfcc100m: The new data in multimedia research.
\newblock {\em Communications of the ACM}, 59(2):64--73, 2016.

\bibitem{li2015you}
Xutao Li, Tuan-Anh~Nguyen Pham, Gao Cong, Quan Yuan, Xiao-Li Li, and Shonali
  Krishnaswamy.
\newblock Where you instagram? associating your instagram photos with points of
  interest.
\newblock In {\em Proceedings of the 24th ACM International on Conference on
  Information and Knowledge Management}, pages 1231--1240, 2015.

\bibitem{liu2014your}
Bo~Liu, Quan Yuan, Gao Cong, and Dong Xu.
\newblock Where your photo is taken: Geolocation prediction for social images.
\newblock {\em Journal of the Association for Information Science and
  Technology}, 65(6):1232--1243, 2014.

\bibitem{wang2017your}
Suhang Wang, Yilin Wang, Jiliang Tang, Kai Shu, Suhas Ranganath, and Huan Liu.
\newblock What your images reveal: Exploiting visual contents for
  point-of-interest recommendation.
\newblock In {\em Proceedings of the 26th international conference on world
  wide web}, pages 391--400, 2017.

\bibitem{bingham2019pyro}
Eli Bingham, Jonathan~P. Chen, Martin Jankowiak, Fritz Obermeyer, Neeraj
  Pradhan, Theofanis Karaletsos, Rohit Singh, Paul~A. Szerlip, Paul Horsfall,
  and Noah~D. Goodman.
\newblock Pyro: Deep universal probabilistic programming.
\newblock {\em J. Mach. Learn. Res.}, 20:28:1--28:6, 2019.

\bibitem{kingma2014adam}
Diederik~P Kingma and Jimmy Ba.
\newblock Adam: A method for stochastic optimization.
\newblock {\em arXiv preprint arXiv:1412.6980}, 2014.

\bibitem{patterson2014sun}
Genevieve Patterson, Chen Xu, Hang Su, and James Hays.
\newblock The sun attribute database: Beyond categories for deeper scene
  understanding.
\newblock {\em International Journal of Computer Vision}, 108(1-2):59--81,
  2014.

\bibitem{griffiths2004finding}
Thomas~L Griffiths and Mark Steyvers.
\newblock Finding scientific topics.
\newblock {\em Proceedings of the National academy of Sciences}, 101(suppl
  1):5228--5235, 2004.

\bibitem{Dacrema2019diversity}
Maurizio~Ferrari Dacrema, Paolo Cremonesi, and Dietmar Jannach.
\newblock Are we really making much progress? a worrying analysis of recent
  neural recommendation approaches.
\newblock In {\em Proceedings of the 13th ACM Conference on Recommender
  Systems}, RecSys '19, page 101–109, New York, NY, USA, 2019. Association
  for Computing Machinery.

\bibitem{gastwirth1972gini}
Joseph~L Gastwirth.
\newblock The estimation of the lorenz curve and gini index.
\newblock {\em The review of economics and statistics}, pages 306--316, 1972.

\end{thebibliography}

\end{document}